\documentclass[final,12pt]{elsarticle}

\usepackage{amssymb}
\usepackage{amsmath}
\usepackage{mathtools}
\newcommand{\notimplies}{%
  \mathrel{{\ooalign{\hidewidth$\not\phantom{=}$\hidewidth\cr$\implies$}}}}
\usepackage{color}
\usepackage{url}
\usepackage{subcaption}
\usepackage{tabularray}
\usepackage{comment}
\usepackage{array} 
\usepackage{graphicx} 

\begin{document}

\begin{frontmatter}

\title{Feature Attribution in 5G Intrusion Detection: A Statistical vs. Logic-Based Comparison}

\author[label1]{Federica Uccello}
\ead{federica.uccello@liu.se}
\author[label1]{Simin Nadjm-Tehrani}
\ead{simin.nadjm-tehrani@liu.se}


\affiliation[label1]{organization={Department of Computer and Information Science, Linköping University},
            city={Linköping},
            country={Sweden}}

\begin{abstract}
With the rise of fifth-generation (5G) networks in critical applications, it is urgent to move from detection of malicious activity to systems capable of providing a reliable verdict suitable for mitigation.
In this regard, understanding and interpreting machine learning (ML) models' security alerts is crucial for enabling actionable incident response orchestration.
Explainable Artificial Intelligence (XAI) techniques are expected to enhance trust by providing insights into why alerts are raised.
Under the umbrella of XAI, interpretability of outcomes is crucially dependent on understanding the influence of specific inputs, referred to as feature attribution.
{A dominant approach to feature attribution statistically associates feature sets that can be correlated to a given alert. This paper investigates its merits against the backdrop of criticism from recent literature, in comparison with feature attribution based on logic.
We extensively study two methods, SHAP and VoTE-XAI, as representatives of each feature attribution approach by analyzing their interpretations of alerts generated by an XGBoost model across three 5G-relevant datasets (5G-NIDD, MSA, and PFCP) covering multiple attack scenarios.
We identify three metrics for assessing explanations: sparsity, how concise they are; stability, how consistent they are across samples from the same attack type; and efficiency, how fast an explanation is generated. 
Our results reveal that logic-based attributions are consistently more sparse and stable across alerts. More importantly, we found a significant divergence between features selected by SHAP and VoTE-XAI.
However, none of the top-ranked features selected by SHAP were missed
by VoTE-XAI. Finally, we analyze the efficiency of both methods, discussing their suitability for real-time security monitoring even in high-dimensional 5G environments (478 features).}
\end{abstract}


\end{frontmatter}

\section{Introduction}
As fifth-generation (5G) networks are deployed in more critical applications, the demand for advanced security mechanisms capable of operating in highly dynamic environments has become more urgent. These mechanisms must be able to detect threats rapidly and respond effectively to evolving attack vectors \cite{benlloch2023distributed}.

Machine Learning (ML) is emerging as a candidate to address complex problems across diverse domains, including healthcare, telecommunications, and autonomous driving \cite{apruzzese2023role}. In recent years, the literature has increasingly explored the potential of applying ML techniques in 5G networks, but has also highlighted some limitations~\cite{morocho2019machine,garrido2023experimental,lilhore2024cognitive}. Although the integration of ML into security applications (such as threat intelligence and malware analysis) has shown some potential, in the area of intrusion detection, some challenges still need to be resolved \cite{lin2023protocol,apruzzese2023role}. 

As in other domains, applying ML to security in 5G settings introduces a range of technical and practical challenges, and a gap remains between theoretical research and practical deployment \cite{arp2024pitfalls}. 

The opaque, ``black-box" nature of many ML models presents a significant obstacle in environments defined as high-risk under the 
European Union’s Artificial Intelligence (AI) Act\footnote{\url{https://digital-strategy.ec.europa.eu/en/policies/regulatory-framework-ai}}, including financial services, healthcare systems, and government infrastructures, many of which may be connected through wireless mobile interfaces. 

Regulatory and standardization bodies such as ENISA and NIST emphasize the need for transparency, accountability, and operational readiness in AI-driven security systems \cite{ENISA,nist}. Yet, these principles are rarely addressed along with performance and explainability trade-offs. Explainable Artificial Intelligence (XAI) is essential for exploiting the potential of  ML model decision-making and to promote trust by human experts \cite{hassija2024interpreting,saeed2023explainable}. Interpretability, the ability to explain or justify model predictions in a human-understandable way \cite{papernot2018deep}, is increasingly recognized as crucial in such contexts. 

In this regard, tree ensembles offer a practical alternative to complex, black-box models due to their simplicity and clarity in structuring decision paths for classification and regression \cite{shin2024fully}. 
According to Chander et al. \cite{chander2025toward}, decision trees can be employed when the features present in a dataset are dependent on each other, as in tabular data, which is the input type to network intrusion detection systems. Moreover, tree ensembles have demonstrated superior predictive performance over neural networks in several real-world, tabular-data applications \cite{shwartz2022tabular}.

In 5G networks, integrating XAI into monitoring systems can help ensure both high accuracy in threat detection and clarity in response generation \cite{zou2024using}. Further research is essential to embed models in orchestrators that maintain interpretability without compromising performance aiming at regulation-compliant AI in 5G environments \cite{5gSurvey}.

This paper addresses the trustworthiness of alarms created based on tabular data within a security monitoring context for critical infrastructure. It aims to provide a deeper understanding of the merits and limitations of representative methods for interpretability within 5G networks. 
Note that the current paper does not aim to advance the state of intrusion detection in 5G networks, but assumes that further research can overcome some of the pitfalls of AI-based techniques for understanding attacks and anomalies \cite{arp2024pitfalls}.

Several techniques are being explored under the umbrella of explainability, including  SHapley Additive exPlanations (SHAP) \cite{lundberg2017shap}, Local Interpretable Model-Agnostic Explanations (LIME) \cite{ribeiro2016should}, Anchors \cite{ribeiro2018anchors}, Counterfactual (CF) \cite{wachter2017counterfactual}, Integrated Gradient (IG) \cite{sundararajan2017axiomatic}, MaxSAT \cite{ignatiev2022using}, and VoTE-XAI \cite{tornblom2025finding}. Multiple surveys cover the applicability of these methods in various domains and ML models \cite{10854503,ouifak2025comprehensive,schwalbe2024comprehensive, mersha2024explainable}. The large majority of the papers referred to in these surveys target classification tasks on images or text data. Since our aim is not to conduct a survey, we have selected representative statistical and logic-based approaches, suitable for tabular data. 
Furthermore, we have selected methods based on feature attribution. These are particularly effective in applications where understanding the influence of specific inputs is crucial \cite{ouifak2025comprehensive}.

Hence, we have excluded those methods whose interpretability goals do not align with feature attribution, such as CF (changes in feature values required to flip the model output) and Anchors (that generates decision rules). We have excluded IG since we restrict our study to non-differentiable ML models. While IG implementations for non-differentiable models exist \cite{Anderson2023}, they are based on approximations, which might hinder the fairness of the comparison. We have excluded LIME since comparative analysis with SHAP reveals that the latter exhibits better quality according to the parameters defined by the authors \cite{schulte2024studying}. Finally, we have excluded MaxSAT, since in comparison to VoTE-XAI against standard AI benchmarks, it has shown slower runtime behavior \cite{tornblom2025finding}. 

Existing statistical techniques, notably SHAP, have become a de facto standard for model interpretability by quantifying the contribution of each feature present in a prediction. However, recent studies \cite{marques2024logic,huang2024failings,notagame} highlight potential shortcomings in the theoretical foundation of statistical methods. We are interested in understanding whether SHAP’s reliance on approximate Shapley values can impact the capture of meaningful relationships between input samples and features appearing in alerts.  In cybersecurity, inaccuracies are critical, as flawed attributions may lead to missed causes, or delayed, ineffective, and even harmful responses.

Logic-based methods ~\cite{marques2024logic}, and in particular those operating on tree ensembles, such as VoTE-XAI, have been proposed to provide minimal, provably correct explanations. Since logic-based explanations are known to have polynomial-time complexity in the size of the model \cite{bassan2023towards,marques2023logic}, it is interesting to assess whether they are useful in a high-dimensional domain like 5G networking. Given that network outages have been shown to last over a day\footnote{https://www.vodafone.pt/en/press-releases/2022/2/cyberattack-on-vodafone-portugal.html}, we believe that studying the capabilities of such methods towards rapid root-cause identification and service restoration is worth attention. Yet, the potential of logic-based approaches in live incident response remains largely unexplored.  

The objective of this study is to evaluate the merits of the selected feature attribution methods in 5G security monitoring. To enable comparison, we focus on three metrics (described in Section \ref{sec:experiments}) that reflect operational requirements. Sparsity \cite{warnecke2020evaluating} captures the conciseness of attribution, important for interpretability in high-dimensional settings and further mitigation actions. Stability \cite{alvarez2018towards} measures the consistency of attribution across alerts of the \emph{same} attack type, useful to assess whether similar inputs yield similar outputs. Efficiency \cite{warnecke2020evaluating} reflects the computational cost of generating attributions and their suitability for real-time security workflows.

The contributions of the paper are as follows:
\begin{enumerate}
    \item We thoroughly analyze the merits of logic-based and statistical interpretations of alerts in a 5G setting based on feature attribution. The evaluation is carried out with data from three different testbeds, focusing on {sparsity} and stability across multiple samples.
    \item We identify the degree of alignment between statistical and logic-based attributions in terms of attack-specific features.
     \item We measure efficiency under two complementary perspectives: a single explanation in real-time contexts, and generating \emph{all} logic-based explanations, potentially relevant in forensic analysis.
\end{enumerate}

The remainder of the paper is organized as follows: Section \ref{sec:related} reviews the related work; Section \ref{sec:background} provides an overview of {XAI and} the feature attribution methods studied in the paper; Section \ref{sec:method} shows the proposed method, while Section \ref{sec:implementation} provides details regarding the use cases. The evaluation criteria are shown in Section \ref{sec:experiments}, and the results are detailed in Section \ref{sec:results}. The outcomes, threats to validity, and future directions are discussed in Section \ref{sec:discussion}. 
Finally, Section \ref{sec:conclusion} ends the paper with closing remarks.

\section{Related Work}
\label{sec:related}
A large body of existing studies on XAI focuses on statistical feature attribution, particularly SHAP and LIME, with limited exploration of logic-based methods in cybersecurity. In this section, we summarize relevant research and position our study within this landscape.
Note that in the following, we use the term "XAI" to refer to a plethora of methods for \emph{interpretability} that are \emph{not} based on feature attribution. Similarly, we will use the word "explanation" to refer to the output of XAI techniques.

\subsection{Explainable AI for Intrusion Detection}
Various approaches have been applied to intrusion detection systems (IDS) and security monitoring to enhance interpretability and improve trust in ML-based threat detection. These studies demonstrate the practical value of applying XAI to IDS. However, in these works, the XAI methods themselves are not the primary object of analysis.

Gaspar et al. \cite{shap_lime} explore SHAP and LIME to interpret predictions from  Multi-Layer Perceptron (MLP)-based IDSs. Their work aims to evaluate the impact of feature attribution on the accuracy of the classifier. A perturbation analysis was conducted by removing or replacing features based on their importance based on SHAP/LIME and observing changes in classification. The study found that, in certain scenarios, relying on the methods negatively impacted the model’s performance, suggesting potential inconsistencies in the attributions. While this work uncovers possible limitations of LIME and SHAP, it neither explicitly highlights nor addresses them. Complementarily, our work attempts to gain a deeper understanding of the consequences of statistical and logic-based attribution.

Gyawali et al. \cite{10620966} combine  SHAP, LIME, and CF to enhance interpretability in IoT security.  By employing these techniques, the researchers seek to translate model predictions into actionable strategies for attack mitigation.
While the results are encouraging, the resulting explanations are implicitly treated as ground truth. Before linking attribution to actions, a critical step would be assessing their accuracy with respect to the data and the context, which is the aim of the present work.

Kalutharage et al. \cite{kalutharage2025neurosymbolic} propose a neurosymbolic approach to IoT security, integrating autoencoder-based anomaly detection with SHAP explanations and domain knowledge encoded in a knowledge graph. {Similarly,} Choi et al. \cite{choi2025attack} develop an attack-specific feature analysis framework that leverages SHAP to optimize feature selection for unsupervised models and autoencoders.
In both works, attribution is used to refine detection mechanisms or improve alert generation \emph{prior to} or \emph{during} model deployment. In contrast, our work focuses on interpreting alerts in operational settings, aiming to support post-detection analysis and decision-making.

\subsection{Explainable AI and Model Performance}
Beyond IDS, multiple works leverage XAI as a tool to improve or diagnose the underlying ML models.

Najibi et al. \cite{najibi2025towards} apply SHAP and LIME to extract the most influential features from deep neural network classifications and use domain knowledge to identify features considered more resistant to evasion. The models are then retrained using only these selected features, resulting in degraded detection performance. Vo et al. \cite{vo2025mdob} similarly use SHAP for dimensionality reduction across various ML models by retraining classifiers on top-ranked features, observing performance degradation in some cases and slight improvements in others. Both studies show that attribution-guided feature selection led to performance degradation in certain scenarios, but the explicit investigation of this behavior was not part of the scope. In contrast, our work aims at investigating feature attribution per se.

Drichel et al. \cite{drichel2023false} used white-box, gradient-based techniques from the iNNvestigate library\footnote{\url{https://github.com/albermax/innvestigate}} to analyze deep learning models trained to reveal botnet activity through Domain Generation Algorithm (DGA) detection. Their goal is to use XAI to uncover bias in the classifier, showing a degradation of the model's performance once the bias is removed. While their findings highlight the value of such techniques for understanding classifier behavior, 
the study uses them as diagnostic tools for model analysis. We apply feature attribution methods with a different scope, assessing their merits for post-detection interpretation.

\subsection{Explainable AI for 5G and Beyond}

Recent works have explored the integration of XAI into 5G/6G architectures, motivated by operational constraints such as scalability, latency, and interpretability of network-level decisions.

Basaran et al. \cite{basaran2024xainomaly} present XAInomaly, an anomaly detection framework for 6G networks. They propose an extension of SHAP designed to operate within the O-RAN architecture. Their work focuses on attribution for intelligence orchestration and network operation, emphasizing real-time feasibility through runtime and resource usage assessment.
Terra et al. \cite{terra2020explainability} apply global and local techniques to analyze SLA violations in 5G network slicing, validating their results with causal inference tools. While a diverse set of XAI techniques is applied, the research does not explore logic-based approaches. 
In contrast to both works, our study explicitly compares statistical and logic-based feature attributions in a security monitoring setting rather than network control.

\subsection{Studies on Feature Attribution}
Several works have examined statistical methods, particularly SHAP and LIME, in networking and security settings. These studies raise concern on conceptual and practical shortcomings of statistical attributions, motivating the need for alternative explanation paradigms.

Fiandrino et al. \cite{fiandrino2023explora} demonstrate that SHAP, when applied out-of-the-box, produces non-intuitive and costly attributions for deep reinforcement learning (DRL) agents in O-RAN systems, as it relies on intermediate representations rather than meaningful inputs. While they propose a graph-based alternative, the scope is limited to network control and does not extend to intrusion detection or broader security monitoring.

Huang and Marques-Silva \cite{huang2024failings} critically analyze SHAP’s shortcomings in security settings, demonstrating that its assumptions of feature independence can lead to misleading attributions in complex datasets. The authors argue for the development of more reliable methods that accurately reflect feature relevance, but proposing an alternative was not in the scope of their work. This highlights the necessity for alternative approaches that do not rely on statistical assumptions.

Marques-Silva \cite{marques2023logic} gave a tutorial in a summer school, highlighting that statistical methods like SHAP and LIME 
often yield inconsistent and logically unsound attributions using small-scale synthetic models. They suggest that abductive and contrastive attribution, computed via Boolean Satisfiability (SAT) and Maximum Satisfiability (MaxSAT), can ensure rigor and minimality in explanations. 

Our work takes this theoretical observation further to evaluate logic-based feature attribution in a practical landscape, applies it to 5G security, and further studies the timeliness of the explanation computations. 
Compared to the above state of the art, we empirically evaluate SHAP and VoTE-XAI based on sparsity, stability, and efficiency in three 5G security use cases. 

Finally, we position our work compared to the work introducing VoTE-XAI. Törnblom et al. \cite{tornblom2025finding} focus on the theoretical underpinnings of the method to derive the explanations in tree ensembles, the soundness and completeness proofs, and some experiments in order to compare the approach to MaxSAT-based approaches. In these experiments, VoTE-XAI showed superior performance. However, the experimental evaluation is limited to common AI benchmarks, with fewer features (ranging from 7 to 60 as the highest dimension). As far as we are aware, the present work is the first to apply logic-based attributions in intrusion detection with a realistic experimental evaluation. 

Considering the wide range of works we have referred to above, we believe that the two representative approaches with their inherent differences, deserve a detailed analysis in realistic case studies on fair grounds. While SHAP and Vote-XAI approaches to attribution are different in nature, we have strived to perform comparative analysis of the usefulness of each on fair grounds.

\section{Background}
\label{sec:background}

In this section, we provide an overview on XAI, followed by concepts of statistical and logic-based feature attribution, with a focus on the methods analyzed in our study.

\subsection{Explainable AI Scope and Concepts}
XAI is an umbrella term which refers to a suite of techniques that aim to enhance transparency, and improve user trust in ML models by adding accountability to their decision-making processes and providing insights into how predictions are generated. The need for explainability was recognized long before the term XAI was introduced in 2016, with early works already highlighting challenges in making model reasoning accessible to end users with precursors presented in a review by Främling \cite{framling2020decision}.

Beyond transparency, XAI aims at other goals, including understanding model behavior, diagnosing errors and biases, enabling informed decision-making, and ensuring that explanations are meaningful for different target audiences. At the same time, open challenges exist, such as lack of rigorous evaluation criteria for explanation quality and the risk of misleading explanations when underlying assumptions are violated \cite{arrieta2020explainable}.

Although there is no general agreement on how explainability should be understood, methods to interpret ML models can be classified along multiple dimensions, including the scope of explanation, the stage of application, and the dependency on model architecture \cite{arrieta2020explainable,schwalbe2024comprehensive}. For instance, global explanations offer insights into the overall decision-making behavior of an ML model, while local explanations focus on individual outputs. Techniques can also be classified based on when they are applied in the ML pipeline: post-hoc methods provide explanations after the model has been trained and deployed. In contrast, ante-hoc refers to building the ability to interpret into a given model. Another classification considers whether a technique depends on a model’s internal structure. Model-agnostic methods explain predictions using only inputs and outputs. Model-specific methods, in contrast, leverage the internal mechanisms of particular ML architectures. 

Within the broad landscape of XAI, feature attribution methods \cite{ouifak2025comprehensive} constitute a class of post-hoc methods, aiming to identify subsets of input features to justify each model prediction.

Table \ref{tab:bg} summarizes the concepts of statistical and logic-based feature attribution, as further detailed in the following subsections.

\begin{table}
\centering
\caption{Conceptual comparison between statistical and logic-based feature attribution.}
\label{tab:bg}
\resizebox{\textwidth}{!}{%
\begin{tabular}{|l|l|l|} 
\hline
\textbf{Aspect}                                                 & \textbf{Statistical Attribution}                                                                                                                       & \textbf{Logic-based Attribution}                                                                                                \\ 
\hline
\begin{tabular}[c]{@{}l@{}}Interpretability \\goal\end{tabular} & \begin{tabular}[c]{@{}l@{}}Which features influence the \\model output, and by how \\much?\end{tabular}                                                  & \begin{tabular}[c]{@{}l@{}}Which features and values justify \\the model output?\end{tabular}                                   \\ 
\hline
Definition of explanation                                       & \begin{tabular}[c]{@{}l@{}}Numerical feature importance \\scores to estimate the \\contribution of individual \\features to the predictions\end{tabular} & \begin{tabular}[c]{@{}l@{}}Minimal sets of feature–value \\pairings sufficient to justify the \\given prediction\end{tabular}  \\ 
\hline
Scope                                                           & \begin{tabular}[c]{@{}l@{}}Local explanations, can be \\aggregated to derive global \\trends.\end{tabular}                                               & \begin{tabular}[c]{@{}l@{}}Local, tied to individual model \\predictions\end{tabular}                                          \\ 
\hline
\begin{tabular}[c]{@{}l@{}}Model \\dependency\end{tabular}      & \begin{tabular}[c]{@{}l@{}}Model-agnostic, with \\optimized variants for specific \\architectures\end{tabular}                                          & \begin{tabular}[c]{@{}l@{}}Model-specific, requiring access \\to the model internals\end{tabular}                              \\ 
\hline
Output                                                          & \begin{tabular}[c]{@{}l@{}}Single attribution vector per \\prediction~\end{tabular}                                                                                                            & \begin{tabular}[c]{@{}l@{}}Single or multiple attributions per \\prediction. Multiple attributions \\ can be compared to each other if \\preferences (domain-based weights \\for features) exist\end{tabular}                                       \\
\hline
\end{tabular}
}
\end{table}

\subsection{Statistical Feature Attribution and SHAP} 

Statistical feature attribution methods form a prominent class of post-hoc techniques. Representative approaches include additive methods such as SHAP and LIME \cite{lundberg2017shap,ribeiro2016should}. Comprehensive surveys report that these methods are widely applied across domains and model classes, with a strong emphasis on classification tasks involving image and text data \cite{10854503,ouifak2025comprehensive,schwalbe2024comprehensive,mersha2024explainable}.

The de facto standard, SHAP, is a technique grounded in cooperative game theory. It explains individual predictions by assigning importance scores, known as Shapley values, to input features. These scores represent the contribution of each feature to the model’s output. 

The Shapley value \( \phi_i \), for a feature \( i \) is defined as shown in Equation \ref{eq:shap_value} \cite{lundberg2017shap}:
\begin{equation} 
\phi_i = \sum_{S \subseteq F \setminus {\left\{i\right\}}} \frac{|S|! , (|F| - |S| - 1)!}{|F|!} \big[ f_{S \cup {\left\{i\right\}}}(x_{S \cup {\left\{i\right\}}}) - f_S(x_S) \big]  
\label{eq:shap_value} 
\end{equation}

Where \( F \) is the set of all features, \( S \subseteq F \setminus \{i\} \) is a subset of features excluding \( i \), \( f_{S \cup \{i\}} \) is the model trained with features in \( S \cup \{i\} \), while \( f_S \) is the model trained with only features in \( S \).  Finally, \( x_{S \cup \{i\}} \) and \( x_S \) are the input values drawn from the respective feature subsets.

SHAP is generally model-agnostic, as it can be applied to a wide range of ML models. However, optimized implementations exist for specific architectures, such as the TreeExplainer\footnote{\url{https://shap.readthedocs.io/en/latest/generated/shap.TreeExplainer.html}} module for tree ensembles.
This method is widely adopted due to its scalability and ability to provide both global and local attributions of model behavior.

SHAP attributions are typically interpreted through visualizations to convey feature contributions at different levels of granularity. Local attributions are commonly shown using force plots or waterfall plots, which illustrate how individual feature contributions combine to produce a specific output. Global attribution is often summarized using beeswarm plots, which aggregate local attributions across many instances to highlight overall feature importance and variability.

\subsection{Logic-based Feature Attribution and VoTE-XAI}
Logic-based (or formal) feature attribution methods aim at computing explanations that are logically sound with respect to a model’s prediction. These approaches typically rely on Boolean reasoning, satisfiability (SAT), or optimization variants such as MaxSAT \cite{ignatiev2022using}\cite{marques2024logic} or abstract interpretations \cite{tornblom2025finding}. Within this category, attributions are defined in terms of feature-value pairs that are sufficient to entail a prediction, rather than estimated contribution scores.

VoTE-XAI \cite{tornblom2025finding} focuses on generating explanations for predictions made by tree ensemble models. The approach provides explanations that are provably correct and minimal with respect to a given prediction inferred from the model.

Specifically, an explanation \( E \) is valid for a prediction \( f(c_1, \dots, c_n) \mapsto d \) if:

\begin{equation}
\bigwedge_{(x_i, c_i) \in E} (x_i = c_i) \implies f(x_1, \dots, x_n) = d
\label{eq:valid_explanation}
\end{equation}

Where \( x_1, \dots, x_n \) are the features, \( c_1, \dots, c_n \) are their respective values, and \( d \) is the model’s prediction \cite{tornblom2025finding}.
An explanation \( E \) is minimal if removing any element from a valid explanation \( E \) invalidates it. The formal definition is provided in Equation \ref{eq:minimal_explanation}:

\begin{equation}
\forall A \subset E, \bigwedge_{(x_i, c_i) \in A} (x_i = c_i) \notimplies f(x_1, \dots, x_n) = d
\label{eq:minimal_explanation}
\end{equation}

Equations \ref{eq:valid_explanation} and \ref{eq:minimal_explanation} have been defined in \cite{tornblom2025finding}.
VoTE-XAI starts with a full explanation consisting of all feature-value pairs involved in a prediction. It then applies a deletion filter that iteratively removes features. An abstract interpretation-based oracle efficiently checks whether each reduced explanation still leads to the same prediction. If so, the feature is discarded. Otherwise, it is kept in the minimal set.

\section{Method}
\label{sec:method}

The objective of this study is to get a deeper understanding of the effectiveness of the above two representative methods in the context of 5G security monitoring. We are interested in assessing whether the two methods perform similarly in terms of identified features in each attack scenario and across similar ones. In addition, we want to understand whether the timing performance of the two methods is appropriate in this 5G context. To be as fair as possible in the comparative analysis, we will use an efficient tree-specific version of SHAP in Section \ref{sub:workflow}, Step 3.

\subsection{Research Questions}
To systematically compare the two methods, we define the following research questions:

\begin{itemize}
    \item RQ1: How do SHAP and VoTE-XAI differ in identifying and prioritizing critical features for a given alert in 5G security monitoring?
    \item RQ2: How efficient are SHAP and VoTE-XAI in generating attributions, and to what extent does this impact real-time response and forensic analysis?

\end{itemize}

\subsection{Investigation Approach}
\label{sub:workflow}
The method is summarized in Figure \ref{fig:workflow}. It consists of four 
steps which are further detailed in Section \ref{sec:experiments}:

\begin{figure}
    \centering
    \includegraphics[width=0.6\textwidth]{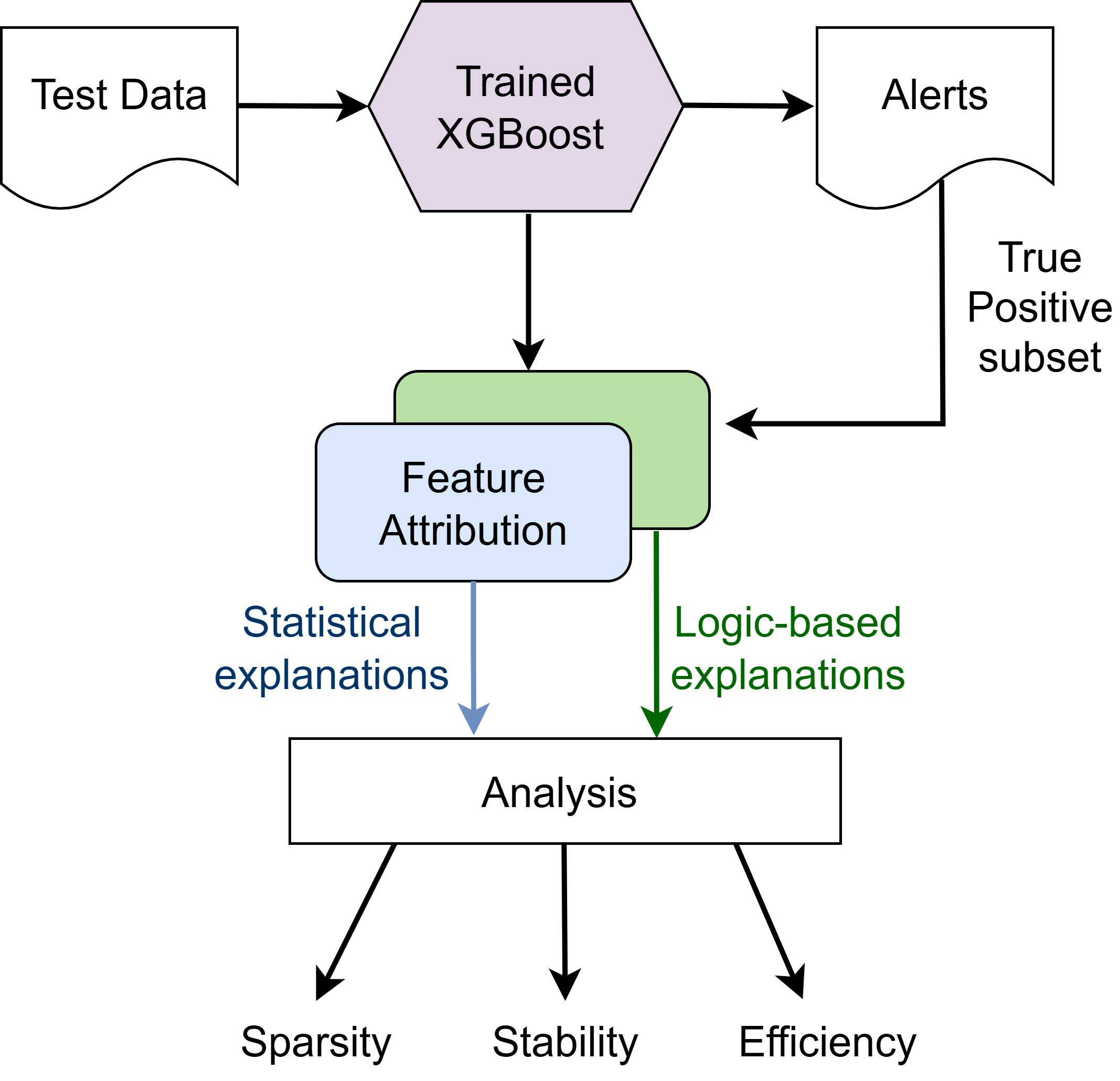}
    \caption{High-level representation of the investigation approach.}
    \label{fig:workflow}
\end{figure}

\begin{enumerate}
    \item \textbf{Raising Alerts with Trained Model}: An XGBoost classifier is trained and used to generate alerts on the test data. The model's classification output serves as the input for subsequent steps.
    \item \textbf{Selection of Alerts for Explanation}: The outputs of the model will be of two categories (negatives and positives). The True Negatives (TNs) are the huge majority of cases in a real scenario, and there is no reason nor way to explain them. They simply pass the monitoring quietly. False Negatives (FNs) correspond to missed threats, and the operator would not be able to act upon something that is not even brought to their attention. Among the positives, the True Positives (TPs) are the most urgent to act upon, and the main reason for attribution, hence the focus of our method. With respect to False Positives (FPs), attribution may be useful to decide to de-prioritize them \cite{colaco2024fast}. The overarching goal of our work is to see how attribution may be useful for future mitigation actions. FPs cannot be mitigated, only identified and ignored. Hence, this Step will focus on TPs\footnote{{However, a comment about how the few FPs in our experiments would be treated when explained is included in section \ref{sub:alert}. Note that the number of such alerts is so small that a statistical evaluation would not be meaningful.}}.
    \item \textbf{Explanation Generation via Feature Attribution}: For each alert, statistical attributions are derived using SHAP TreeExplainer optimization, and logic-based ones are computed using VoTE-XAI. At this stage, the feature attribution engines take as input both the selected alerts and the XGBoost model. This is necessary since the methods need to access the internal structure of the model to have the context within which alerts are raised.
    \item \textbf{Analysis}: The resulting attributions are evaluated based on the metrics defined in Section \ref{sec:experiments}. The findings help evaluate the sparsity and stability of the attributions, determine the degree of alignment between the two methods, and assess computational efficiency.
\end{enumerate}

\section{Incident-response Case Studies}
\label{sec:implementation}
Open datasets for studying 5G network attacks and defenses are few and far between. The existing ones are not ideal due to a severe imbalance in data (benign/attack) categories since they are created in research labs for attack classification or the study of a specific mitigation response. They contain far more attacks than benign data compared to real world.
However, our goal here is to create interpretations of alerts raised while monitoring, attributing the attack to identified feature sets.
Therefore, potential deficiencies in the datasets will not be relevant for our work, which focuses on the alert level. 

This section describes the datasets used and the ML models created for detection of attacks, their training configurations, and performance.

\subsection{Dataset Description}

We use three available datasets for studying 5G attacks: the 5G Network Intrusion Detection (5G-NIDD) \cite{5gnidd}, Multistep Attack (MSA) \cite{samin2025gotta}, and Packet Forwarding Control Protocol (PFCP) \cite{amponis2022threatening} datasets.
Although we could have applied our approach to better-known but not 5G-related network intrusion detection benchmarks, we observe that those datasets are rather low-dimensional relative to the complexities in 5G networks reflected in terms of the number of features. For instance, CICIoT2023 \cite{neto2023ciciot2023} and CIC UNSW-NB15 \cite{mohammadian2024poisoning} contain 46 and 76 features, respectively, which is fewer than even our use case with the lowest dimension in the next subsections.

All selected datasets have been preprocessed by removing duplicate entries and encoding labels using \textit{sklearn LabelEncoder}, following standard ML pipeline procedures. Then, the flows were split into training and test subsets using an 80:20 ratio and preserving the class proportions of the full datasets.

The details of the adopted datasets are in the following, and summarized in Table \ref{tab:datasets}. 

\begin{table}
\centering
\caption{Summary of datasets' dimensionality, class balance, and train/test splits.}
\label{tab:datasets}
\resizebox{\textwidth}{!}{%
\begin{tabular}{|l|c|c|cc|cc|c|} 
\hline
\textbf{Dataset}                                    & \textbf{\# Features} & \textbf{Total Samples} & \multicolumn{2}{c|}{\textbf{Train Set}} & \multicolumn{2}{c|}{\textbf{Test Set}} & \textbf{\# Attack Types}  \\ 
\cline{4-7}
                                                    &                      &                        & \textbf{Benign} & \textbf{Malicious}    & \textbf{Benign} & \textbf{Malicious}   &                           \\ 
\hline
5G-NIDD \cite{5gnidd}              & 92                   & 1,215,890              & 381,904         & 590,812               & 95,833          & 147,341              & 8                         \\ 
\hline
MSA \cite{samin2025gotta}          & 478                  & 25,160                 & 719             & 19,409                & 186             & 4,846                & 21                        \\ 
\hline
PFCP \cite{amponis2022threatening} & 83                   & 7,199                  & 1,028           & 4,114                 & 411             & 1,646                & 4                         \\
\hline
\end{tabular}
}
\end{table}

\subsubsection{5G-NIDD}

The 5G-NIDD provides labeled network traffic data, distinguishing between normal operations and attack scenarios. It includes 92 network flow features, such as flow duration, packet size statistics, protocol types, flow flags, and packet inter-arrival times. The original documentation was not complete in terms of explaining the features of the data set. Thus, additional work was needed to understand the meaning of the feature names referred to. These are useful for understanding the rest of this paper. Table \ref{tab:features} presents a subset of the most relevant network flow features identified across the methods applied in this study.

The dataset supports both binary and multiclass classification for intrusion detection, and it includes two primary attack categories: 
\begin{itemize}
    \item Denial of Service (DoS): Includes volumetric (ICMP Flood, UDP Flood), protocol-based (SYN Flood), and application-layer attacks (HTTP Flood, Slowrate DoS).
    \item Port Scans: Comprising SYN Scan, TCP Connect Scan, and UDP Scan.
\end{itemize}

\begin{table}[htbp]
\centering
\caption{Description of the most meaningful features of the 5G-NIDD data set appearing in attributions from both evaluated methods.}
\resizebox{0.75\textwidth}{!}{%
\begin{tabular}{|l|p{0.75\linewidth}|}
\hline
\textbf{Feature} & \textbf{Description} \\ \hline
cs0 & Quality of Service metric in the IP header (default best-effort service class). \\ \hline
dHops & Number of hops from destination (255 - dTtl). \\ \hline
DstBytes & Total bytes sent from the destination to the source. \\ \hline
DstTCPBase & Destination TCP base sequence number (initial sequence number used by the destination in the TCP connection). \\ \hline
DstWin & Destination TCP window advertisement value. \\ \hline
dTtl & Destination to source Time-to-Live value. \\ \hline
Dur & Total duration of the flow in seconds. \\ \hline
ECO & ICMP Echo Request (Ping request). \\ \hline
FIN & TCP connection termination flag. \\ \hline
INT & Initial transaction state (seen in UDP flows when the first packet is detected). \\ \hline
Load & Total bits per second for the flow (Sload + Dload). \\ \hline
Loss & Total packets lost (sloss + dloss). \\ \hline
Offset & Fragmentation offset value in the IP header. \\ \hline
pLoss & Number of packets lost from source to destination. \\ \hline
Rate & Total bits per second (same as Load). \\ \hline
REQ & TCP connection request (SYN seen, connection attempted). \\ \hline
RST & TCP connection reset (rejected or forcibly closed). \\ \hline
Seq & Flow sequence number. \\ \hline
sHops & Number of hops from the source (255 - sTtl). \\ \hline
sMeanPktSz & Mean size of packets transmitted by the source. \\ \hline
SrcBytes & Total bytes sent from the source to the destination. \\ \hline
SrcLoad & Source bits per second (Sload). \\ \hline
SrcPkts & Number of packets sent from the source to the destination. \\ \hline
SrcRate & Source bits per second (Sload). \\ \hline
SrcTCPBase & Source TCP base sequence number (initial sequence number used by the source). \\ \hline
SrcWin & Source TCP window advertisement value. \\ \hline
Start & Timestamp when the flow started. \\ \hline
sTtl & Source to destination Time-to-Live value. \\ \hline
SynAck & Time between SYN and SYN-ACK during TCP handshake. \\ \hline
TotBytes & Total bytes transferred (SrcBytes + DstBytes). \\ \hline
TotPkts & Total packets transferred (SrcPkts + Dpkts). \\ \hline
udp & Flow uses UDP protocol. \\ \hline
\end{tabular}
}
\label{tab:features}
\end{table}

\subsubsection{MSA Dataset}
The MSA dataset includes labeled data representing a wide variety of attacks on the control plane of 4G/5G networks. 
It was selected to evaluate the scalability of explanation methods, given its high dimensionality in terms of features (478 in total). To help better understand the rest of the paper, the most relevant features identified in the study are explained in Table \ref{tab:features_MSA}.

\begin{table}[htbp]
\centering
\caption{Description of the most meaningful feature groups of the MSA appearing in feature attributions from both methods evaluated in this paper.}
\resizebox{\textwidth}{!}{%
\begin{tabular}{|l|p{0.75\linewidth}|}
\hline
\textbf{Feature Group} & \textbf{Description} \\ \hline
NAS EPS EMM & Handles LTE mobility management: attachment, tracking area update, UE identification, and EMM state transitions. Useful for detecting rogue base stations and tracking anomalies.  
Includes: F6: nas-eps-emm-id-type2-unmaskedvalue, F13: nas-eps-emm-guti-type-size, F16: nas-eps-emm-128eea1-unmaskedvalue, F20: nas-eps-emm-switch-off-size, F21: nas-eps-emm-id-type2-size. \\ \hline
NAS Security & Monitors NAS-level ciphering and integrity algorithms. Essential for identifying weak security configurations.  
Includes: F8: nas-eps-security-header-type-value, F17: nas-eps-msg-auth-code-show. \\ \hline
NAS Identifiers E.212 & Contains global subscriber/operator identifiers. Useful for profiling and roaming analysis.  
Includes: F4: e212\_gummei\_mcc\_size, F10: e212-tai-mnc-value, F23: e212-tai-mnc-show, F25: e212-tai-mnc-unmaskedvalue. \\ \hline
NAS Message Internals Encoding & Represents NAS message structure and encoding behavior. Helps detect malformed messages or optional feature usage.  
Includes: F7: nested-field2-value, F12: nested-field1-value, F14: nested-field1-show. \\ \hline
GSM A / DTAP (2G/3G Signaling) & Legacy 2G/3G signaling for mobility and authentication. Important for inter-RAT handover and fallback analysis.  
Includes: F1: gsm-a-oddevenind-unmaskedvalue, F2: gsm-a-spare-bits-unmaskedvalue, F22: gsm-a-dtap-add-ci-value, F24: gsm-a-extension-size. \\ \hline
ASN.1 Encoding Diagnostics & Captures encoding metadata and diagnostic flags. Important for debugging and feature engineering. \\ \hline
\end{tabular}%
}
\label{tab:features_MSA}
\end{table}

The 21 distinct types of attacks are classified in 4 categories. Some of the attacks are significantly underrepresented in the dataset, which can limit their usefulness for training and evaluation purposes. Therefore, we selected the following subset of attacks to include all the categories:

\begin{itemize}
    \item DoS: Includes several variants, such as ``NAS counter Desynch attack", ``Authentication relay attack", ``Handover hijacking", ``RRC replay attack", ``Incarceration with rrcReestablishReject".
    \item Bidding Down Attacks: These attacks aim to downgrade the network capabilities of the User Equipment (UE) by exploiting vulnerabilities in unprotected signaling messages. The selected attacks include bidding down with AttachReject and with TAUReject. Applying this attack to a set of UEs would amount to a regional outage scenario. 
    \item Location Tracking: Represented by the Location tracking via measurement reports attack, where the attacker leverages the control plane to extract detailed location information from victim UEs.
    \item Battery Drain: These attacks force excessive signaling, state transitions, or repeated cryptographic operations, resulting in rapid energy depletion in (a set of) UEs. 
\end{itemize}
 
\subsubsection{PFCP Dataset}

The focus of this dataset is the security issues of the Packet Forwarding Control Protocol (PFCP), which is used in the N4 interface between the Session Management Function (SMF) and the User Plane Function (UPF) in the 5G core. 
The dataset comprises a total of 83 network features (described in detail in the dataset documentation \cite{amponis2022threatening}) and 7,199 labeled flows.

The malicious flows correspond to the following attack scenarios:
\begin{itemize}
    \item PFCP Session Establishment DoS Attack: Flooding the UPF with legitimate session establishment requests to exhaust its processing capacity.
    \item PFCP Session Deletion Flood: Overwhelming the UPF with session deletion requests, forcing disconnection of UEs from their active data sessions.
    \item PFCP Session Modification Flood (DROP): Injecting modifications to Forwarding Action Rules (FAR) that apply the DROP action, resulting in packet discarding for affected sessions. While this disconnects the UE from the Data Network (DN), the UE remains connected to the 5G core via the Radio Access Network (RAN).
    \item PFCP Session Modification Flood (DUPL): Duplicating session rules to create multiple possible forwarding paths for identical traffic, leading to potential packet duplication and unexpected network behavior. At scale, this attack can cause a distributed DoS across multiple UEs.
\end{itemize}

\subsection{Machine Learning Model}
\label{sub:model}
For classification of the attacks, we tested three different tree ensemble models: XGBoost, Random Forest, and LightGBM, with very similar results in terms of detection performance.

Then, the XGBoost Classifier was selected due to its ability to handle class imbalance effectively using the scale\_pos\_weight parameter and to prevent overfitting via regularization techniques inherent to its algorithm. Additionally, there is evidence of its suitability for network intrusion detection tasks and high scalability \cite{edozie2025artificial}.

Hyperparameter tuning was performed using RandomizedSearchCV from the scikit-learn library.
The model was configured with the hyperparameters outlined in Table \ref{tab:model}, selected based on the results of the RandomizedSearchCV optimization process. The selection of alerts is based on binary classification, but multiclass labels are retained for later analysis.

\begin{table*}[h!]
\centering
\caption{Hyperparameters of the XGBoost Classifiers, selected using the best ones from RandomizedSearchCV.}
\begin{tabular}{|l|l|p{8cm}|}
\hline
\textbf{Hyperparameter} & \textbf{Value} & \textbf{Description} \\ \hline
\texttt{scale\_pos\_weight} & 2.2 & Balances the positive and negative classes to handle class imbalance. \\ \hline
\texttt{n\_estimators} & 100 & The number of boosting rounds (trees) in the model. \\ \hline
\texttt{max\_depth} & 10 & Maximum depth of a tree, controlling model complexity and risk of overfitting. \\ \hline
\texttt{learning\_rate} & 0.35 & Step size shrinkage to prevent overfitting and adjust the weight of new trees. \\ \hline
\end{tabular}

\label{tab:model}
\end{table*}

\section{Experiment Design and Evaluation Strategy}
\label{sec:experiments}
This section {describes in more detail how we applied step 4 of our method from Section \ref{sec:method} in the context of the 5G case studies from Section \ref{sec:implementation}.} The results are shown in Section \ref{sec:results}. 
The source code and data used in this study are available on GitLab \footnote{Source Code: \url{https://gitlab.liu.se/feduc51/5GXAI}.}. All the experiments were executed in a Debian-based virtual machine equipped with an Intel Core Ultra 7 155U CPU and 32 GB RAM.

\subsection{Alert Selection}
\label{sub:alert}

Our experiments took the test sets, selecting malicious samples in a stratified manner across all attack classes (Step 2 in Section \ref{sub:workflow}). That is, we identified the indices of samples classified as malicious and retrieved their corresponding attack class labels. 

Before performing the feature attribution analysis, we need to select TPs among the possible ones. While selecting, we performed stratified sampling to ensure a balanced selection across attack types, allocating an equal number of samples per class, aiming to keep the total TPs per dataset below 100.
This number was selected so that the validation of all explanations manually would be feasible.

More specifically, for each attack class, we first identified all malicious samples belonging to that class and performed stratified sampling without replacement. This means that once a sample is selected, it is not chosen again. However, replacement was incorporated as a fallback mechanism to ensure that if a particular class had fewer samples than required, additional instances could be drawn to maintain balance. 

The final sets adopted for the experiments consisted of 268 across the three datasets: 88 TP samples for the 5G-NIDD dataset (11 per attack class), 80 TP samples for the MSA dataset (8 per attack class), and 100 TP samples for the PFCP dataset (25 per attack class). 

We realize that, to an operator, both TPs and FPs need explanations. As seen in Table \ref{tab:performance}, the number of FPs is very low. However, we examined explanations for all FP cases in the 5G-NIDD dataset. The goal was to find out whether explanations might aid in discerning classification errors in some way.
Details are provided in Section \ref{sub:fp}.

\subsection{Evaluation Metrics}

The field of XAI has seen a proliferation of evaluation metrics, with no universally accepted standard, \cite{pawlicki2024evaluating}.
Hence, to evaluate the performance of both statistical and logic-based approaches, we adopted three metrics: sparsity, stability, and efficiency
, drawing on definitions from Warnecke et al. \cite{warnecke2020evaluating} (for sparsity and efficiency), and from Alvarez-Melis and Jaakkola (\cite{alvarez2018towards} for stability).

\textbf{Sparsity:} An explanation is considered meaningful if it involves only a limited number of features. It can be considered the opposite of verbosity. This type of interpretability measure is intended to reduce the cognitive load for analysts in a human-in-the-loop scenario. Thus, we evaluate how concise the attributions are in highlighting the most relevant features. 

\textbf{Stability:} Stability is a widely discussed but not universally agreed-upon metric in XAI. It can be defined as the consistency of an output to similar input instances. In this study, stability refers to the degree to which attributions remain similar within alert samples from the \emph{same} attack class. If this varies significantly within the same class, it might indicate the inability to generate consistent attributions.

\textbf{Efficiency:} An attribution method is deemed efficient if it can be considered plausible in an operational workflow. 
Real-world threat detection scenarios, such as those in 5G networks, require explanations that are not only accurate and interpretable but also timely (a notion that we will discuss in Section \ref{sub:time}).

Table \ref{tab:metrics_summary} summarizes the chosen metrics, how they are evaluated, and their interpretation.

\begin{table}[htbp]
\centering
\caption{Summary of evaluation metrics and procedures for SHAP and VoTE-XAI.}
\label{tab:metrics_summary}
\resizebox{\textwidth}{!}{%
\begin{tblr}{
  cell{2}{1} = {r=2}{},
  cell{2}{4} = {r=2}{},
  cell{4}{1} = {r=2}{},
  cell{4}{4} = {r=2}{},
  cell{6}{1} = {r=2}{},
  cell{6}{4} = {r=2}{},
  vlines,
  hline{1-2,4,6,8} = {-}{},
  hline{3,5,7} = {2-3}{},
}
\textbf{Metric} & \textbf{Method} & \textbf{Measurement Procedure}                                                                              & \textbf{Interpretation}                                                       \\
Sparsity        & SHAP            & {Number of features with positive \\attribution per sample}                                                 & {Reflects the number of features \\presented to the analyst, independent \\of method}      \\
                & VoTE-XAI        & {Number of features in a single \\minimal explanation per sample}                                           &                                                                                            \\
Stability       & SHAP            & {Fluctuation of min/mean/max \\attribution values per feature \\within each attack class}                   & {Assesses whether similar alerts \\from the same attack class yield \\consistent features} \\
                & VoTE-XAI        & {Frequency of feature occurrence \\across all minimal explanations \\within each attack class}              &                                                                                            \\
Efficiency      & SHAP            & {Runtime to generate one explanation \\per alert}                                                           & {Measures end-to-end computational\\time under identical conditions}                       \\
                & VoTE-XAI        & {Runtime to generate one minimal \\explanation per alert and all minimal \\explanations evaluated separately} &                                                                                            
\end{tblr}
}
\end{table}

\subsection{Analysis of Sparsity and Stability}
Due to the different nature of statistical and logic-based attributions, rather than forcing identical representations, each metric is instantiated to reflect the native output of each method while measuring the same interpretability aspect.

This analysis examines how concise SHAP and VoTE-XAI explanations are (using sparsity) and whether they remain consistent within the same attack class (using stability). We ran all the test data flows from each dataset as mentioned in Section \ref{sec:implementation} with the trained attack classification model. Then, both explanation approaches were applied to the subsets of alerts selected as explained in Section \ref{sub:alert}.

For SHAP, we employ the TreeExplainer module. Explanations are computed by selecting all the features with positive attribution, resulting in one individual explanation per sample. Note that a very low SHAP value in that approach signifies a small likelihood of the relevance of a feature, but we chose to include all such features so that no potential attributions would be missed due to choosing an arbitrary threshold. In the context of our study, features with negative SHAP values are not considered relevant, as they represent factors that did not contribute to the raising of that alert. Hence, the potential usefulness of these falls outside the scope of our analysis. 

In contrast to SHAP, VoTE-XAI provides logic-based, minimal explanations that highlight the logically valid features for a given classification. There may be different minimal explanations for the same alert, but they are all valid explanations. The method allows calculation of \emph{all} minimal explanations or a \emph{single} one for a given alert. 

For SHAP, sparsity is computed as the number of features with positive attribution values per sample. For VoTE-XAI, we compute the number of features in a single minimal explanation per sample. We believe this procedure is fair as it gets, as sparsity directly reflects the number of features that would be presented to the analyst for both methods.

Stability for SHAP is assessed by computing the minimum, maximum, and mean SHAP values per feature. This helps identify how consistently SHAP associates importance to specific features across different samples of the same attack class. It also shows whether certain features dominate explanations or if feature attributions fluctuate widely. By comparing SHAP distributions within each attack class, we assess feature stability (consistent attribution) vs. variability (fluctuating importance across samples from the same attack type).

To assess stability among VoTE-XAI attributions, all minimal explanations were computed for each sample and attack class. The features were ranked based on their occurrence across all minimal explanations, then we identify features that appear consistently across all minimal explanations for each sample of each attack type. For both methods, stability shows whether similar alerts from the same attack class yield consistent features, despite differences in attribution formalism.

\subsection{Divergence Among the Methods}
The degree of alignment between SHAP and VoTE-XAI feature attributions relates to RQ1 in Section \ref{sec:method}, and is investigated by comparing the features according to SHAP's top-ranked feature importance values with the most frequently occurring features in VoTE-XAI minimal explanations for each attack class. 

A high agreement between these methods suggests that SHAP’s statistical attributions align with VoTE-XAI's logic-based explanations. Conversely, low agreement indicates a divergence between the methods. Given that VoTE-XAI provides interpretations that are logically sound with respect to the underlying model, any attribution is a direct consequence of adopting that given model for detection. Whether the attribution is valid (as a ground truth in the security domain) or not is therefore relative to the accuracy of a model.

Whether statistical attributions provide a matching to a security-aligned ground truth is subject to the same uncertainty, too.  However, it is important to compare them against the same model’s inferences to see if any limitations become evident.

\subsection{Efficiency Evaluation}

The purpose of this analysis is to evaluate the efficiency of SHAP and VoTE-XAI (relating to RQ2) by measuring the time required to generate explanations. Efficiency is fairly compared by measuring end-to-end explanation generation time under identical conditions (one single explanation per alert). In addition, we separately evaluate the efficiency of VoTE-XAI in computing all minimal explanations for each alert.

Real-time threat detection in 5G networks requires fast and interpretable explanations to support immediate decision-making. If an explainability method is computationally expensive, it risks becoming impractical for deployment in security operations. However, slower methods can still be useful in forensic contexts.
Our analysis is based on a larger TP alert set consisting of 3600 samples (1200 samples for each dataset), different from the ones used for sparsity, stability and divergence, but selected using the same procedure as described in Section~\ref{sub:alert}. Samples were drawn across attack classes to preserve the class diversity present in the original datasets. The size of this subset was chosen as a compromise between computational cost and coverage of multiple attack classes. The observed trends are expected to scale consistently due to the per-sample nature of feature attribution generation.

\section{Results}
\label{sec:results}
This section presents the experimental results, aiming to answer RQ1 (Subsections \ref{sub:spars}, \ref{sub:stab}, 
and \ref{sub:divergence}) and RQ2 (Subsection \ref{sub:time}) following Steps 3 and 4 of the workflow described in Section \ref{sub:workflow}.

The XGBoost model was evaluated on the selected datasets (Step 1 in the workflow from Section \ref{sub:workflow}). 
As shown in Table \ref{tab:performance}, the metrics used include TN, FN, FP, TP, Accuracy, Precision, Recall, and F1-score\footnote{Definition of these metrics is standard but can be found in the Scikit-learn user guide: \url{https://scikit-learn.org/stable/modules/model_evaluation.html}}. Note that the perfect recall in the PFCP Dataset is likely the consequence of the small size of the original dataset. Although the actual detection quality is not the purpose of our work, we present the outcomes of step 2 for completeness and context.

To get an idea about the set of alerts that can be subject to explanation, 
the number of samples in each category in Table \ref{tab:performance} is useful, only to provide information about the test environment for the explanation study in the next section, in particular the number of available TPs.

In the following, we use the multiclass labels retained from Step 1 (Section \ref{sub:model}) to analyze the attributions for \emph{each} individual attack class.

\begin{table}[ht]
\centering
\caption{Confusion matrix elements and evaluation metrics for the XGBoost Classifier for both datasets.}
\resizebox{\textwidth}{!}{%
\begin{tabular}{|l|c|c|c|c|c|c|c|c|}
\hline
\textbf{Dataset} & \textbf{TN} & \textbf{FN} & \textbf{FP} & \textbf{TP} & \textbf{Accuracy} & \textbf{Precision} & \textbf{Recall} & \textbf{F1-Score} \\ \hline
5G-NIDD & 95796 & 35 & 37 & 147306 & 0.9997 & 0.9997 & 0.9998 & 0.9997 \\ \hline
MSA & 152 & 4 & 34 & 4842 & 0.9924 & 0.9930 & 0.9992 & 0.9951 \\ \hline
PFCP & 409 & 0 & 2 & 1646 & 0.9990 & 0.9988 & 1.0000 & 0.9994 \\ \hline
\end{tabular}%
}
\label{tab:performance}
\end{table}

\subsection{Sparsity Evaluation}
\label{sub:spars}
SHAP assigns an importance score to each feature, indicating its contribution to the model’s output. On average, it is observed that SHAP explanations are consistently {less sparse} than VoTE-XAI's ones across the three use cases.

Specifically, each SHAP explanation includes 20–25 features with positive attributions out of the original 92 in use case 1 (5G-NIDD). For use case 2 (MSA), SHAP explanations identify between 59 and 90 features per sample, among the 478 from the full set. For use case 3 (PFCP), each SHAP explanation includes 16-20 features out of the original 83.

When computing only one minimal explanation for each sample, it is observed that VoTE-XAI provides {sparser} explanations compared to SHAP. For use case 1 (5G-NIDD), 6-15 features were identified on average. In use case 2 (MSA), explanations contain 27–41 features out of the 478 in the original dataset, less than half the size of SHAP’s output. For use case 3 (PFCP), each VoTE-XAI explanation consists of only 4–6 features. 
 
\subsection{Stability Evaluation}
\label{sub:stab}
In all the use cases, SHAP values fluctuate significantly across different samples of the same attack class. 
In contrast, VoTE-XAI provides highly consistent explanations across samples: features identified as critical by VoTE-XAI appear in nearly every explanation for a given attack class. More details are in Tables \ref{tab:SHAP_stab} and \ref{tab:Vote_stab}.

\begin{figure}[htbp]

\centering
\begin{subfigure}[t]{0.45\textwidth}
    \centering
    \includegraphics[width=\linewidth]{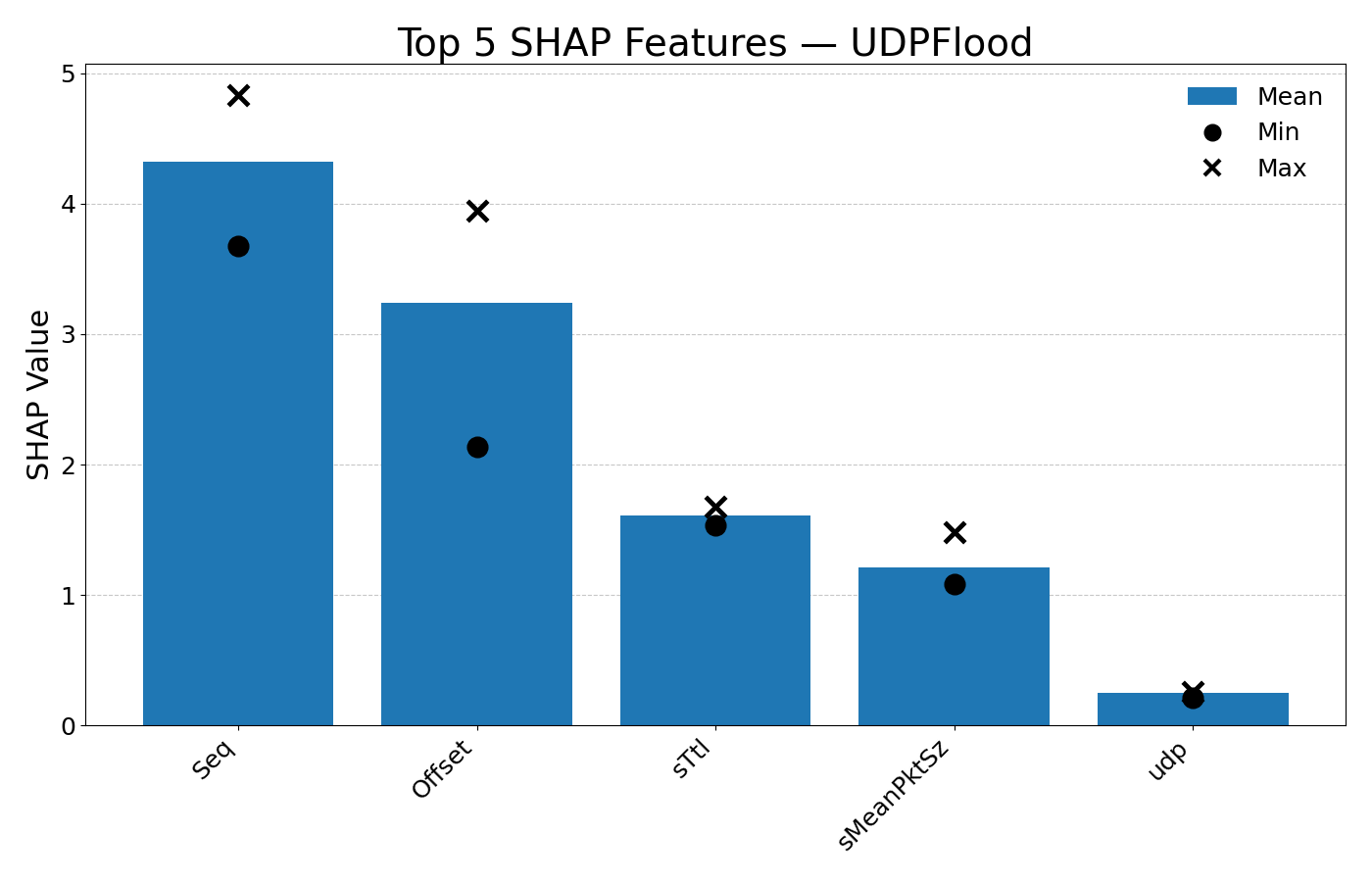}
    \caption{UDPFlood}
    \label{fig:shap_1a}
\end{subfigure}
\hfill
\begin{subfigure}[t]{0.45\textwidth}
    \centering
    \includegraphics[width=\linewidth]{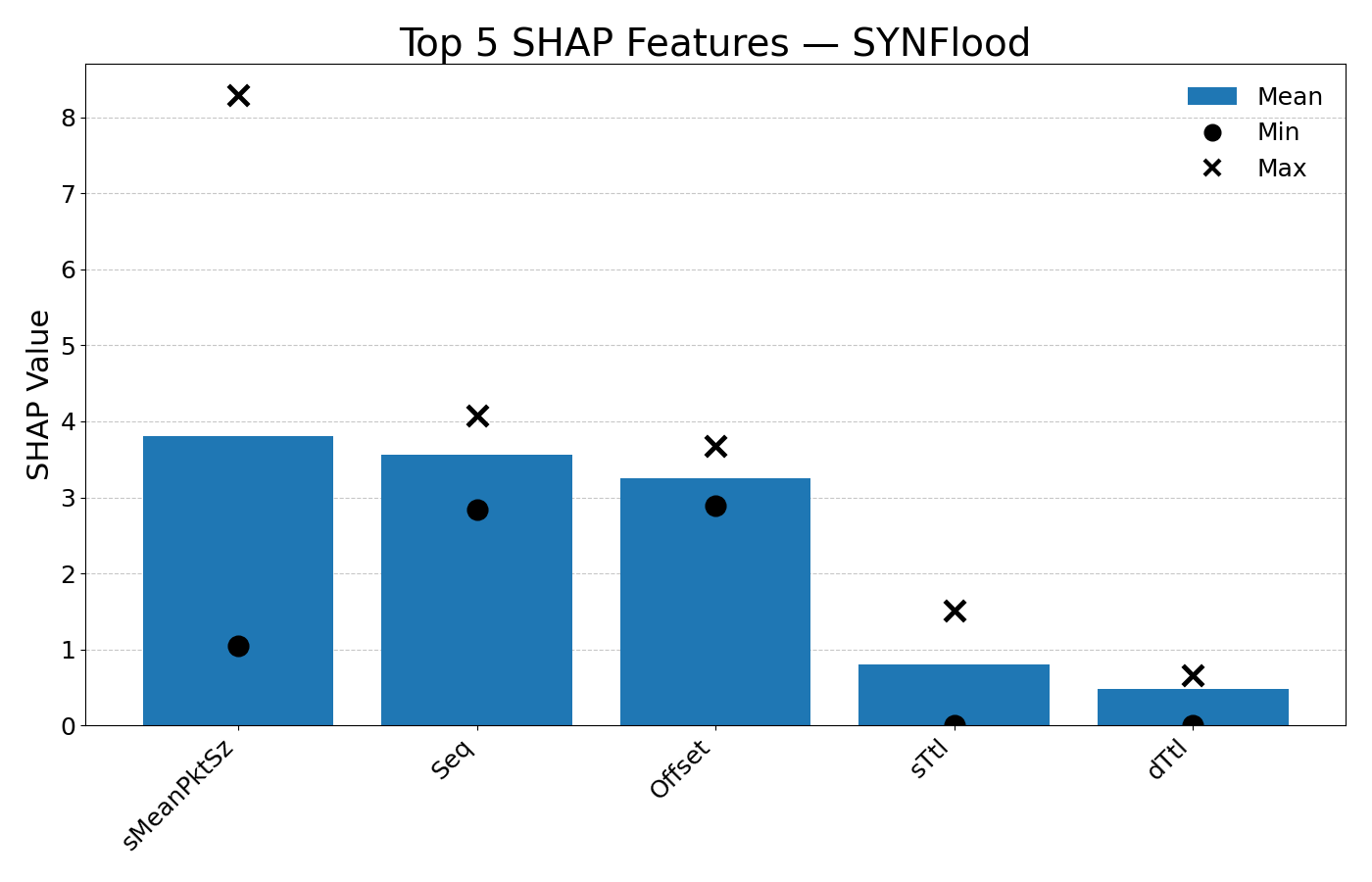}
    \caption{SYNFlood}
    \label{fig:shap_1b}
\end{subfigure}

\vskip\baselineskip
\begin{subfigure}[t]{0.45\textwidth}
    \centering
    \includegraphics[width=\linewidth]{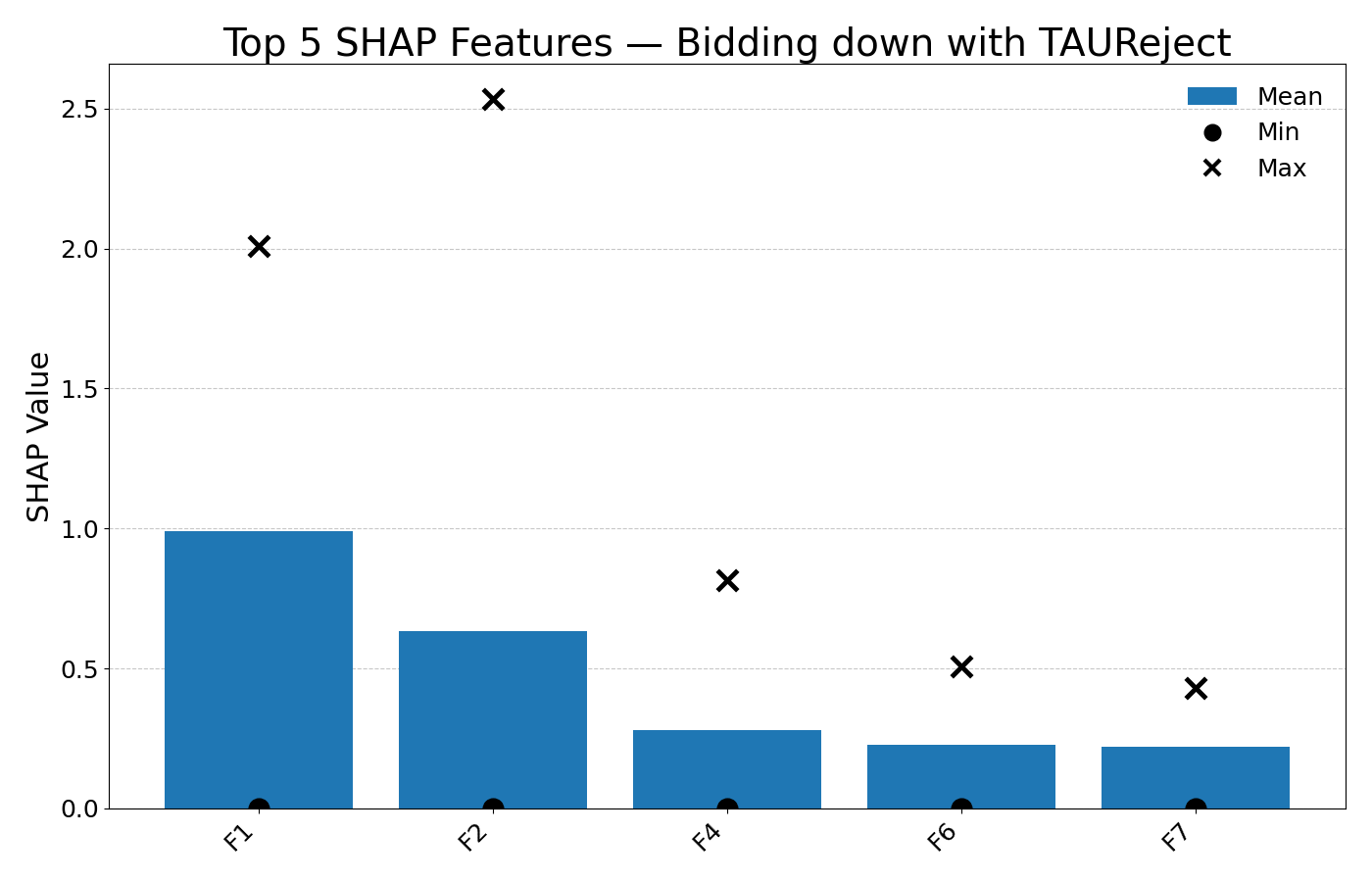}
    \caption{Bidding down with TAUReject}
    \label{fig:shap_2a}
\end{subfigure}
\hfill
\begin{subfigure}[t]{0.45\textwidth}
    \centering
    \includegraphics[width=\linewidth]{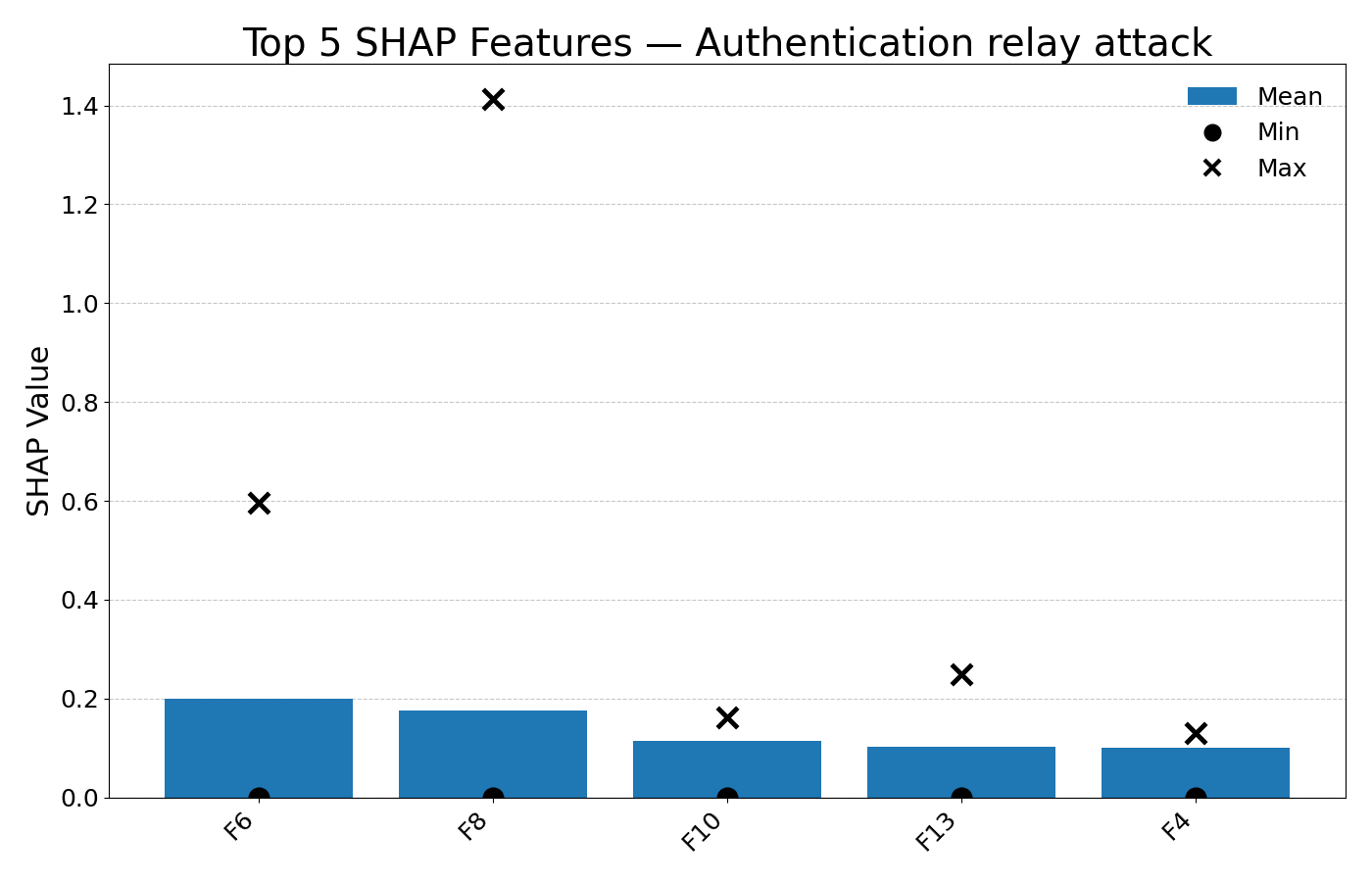}
    \caption{Authentication relay attack}
    \label{fig:shap_2b}
\end{subfigure}    
\vskip\baselineskip
\begin{subfigure}[t]{0.45\textwidth}
    \centering
    \includegraphics[width=\linewidth]{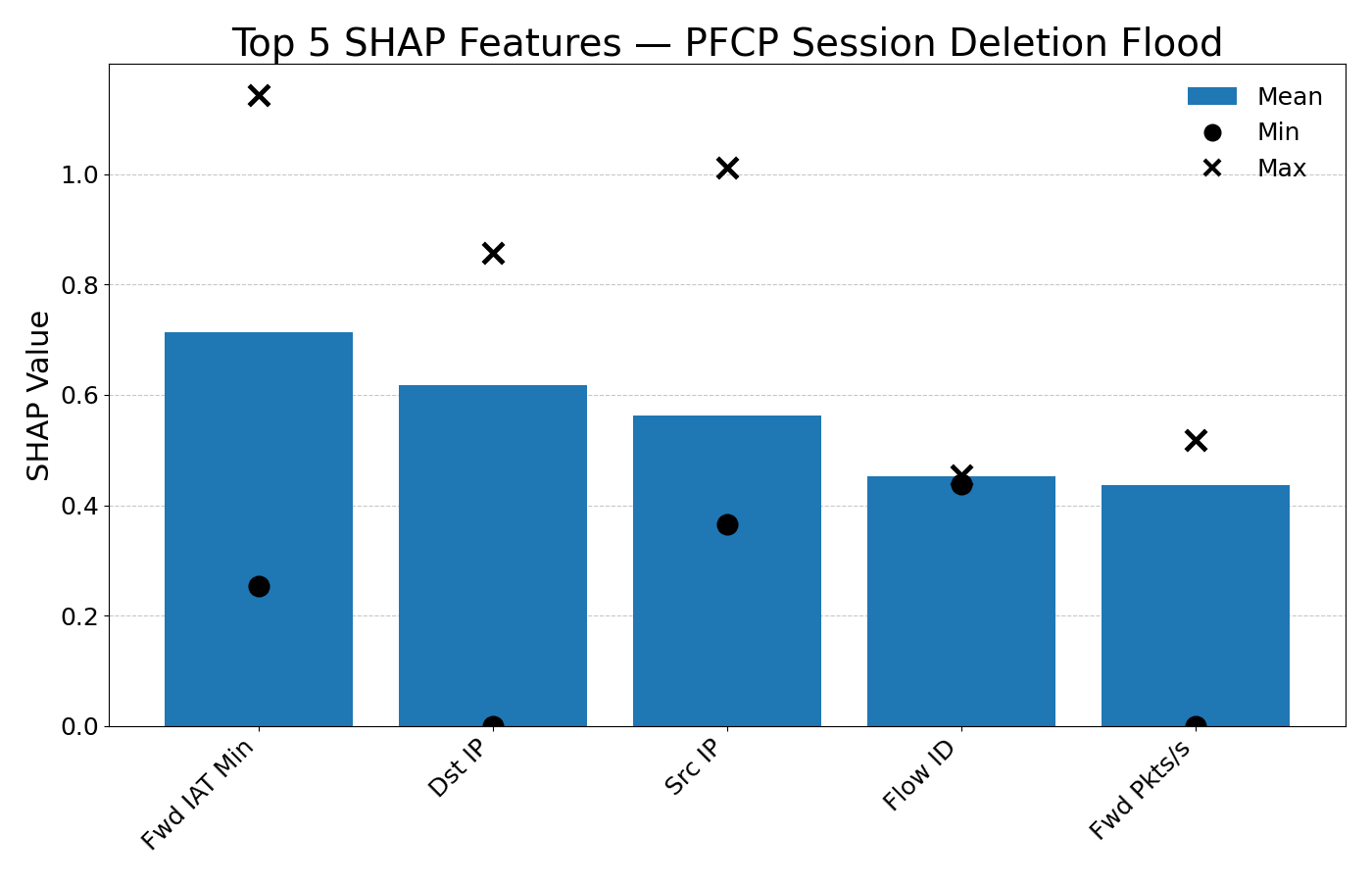}
    \caption{PFCP Session Deletion Flood}
    \label{fig:shap_3a}
\end{subfigure}
\hfill
\begin{subfigure}[t]{0.45\textwidth}
    \centering
    \includegraphics[width=\linewidth]{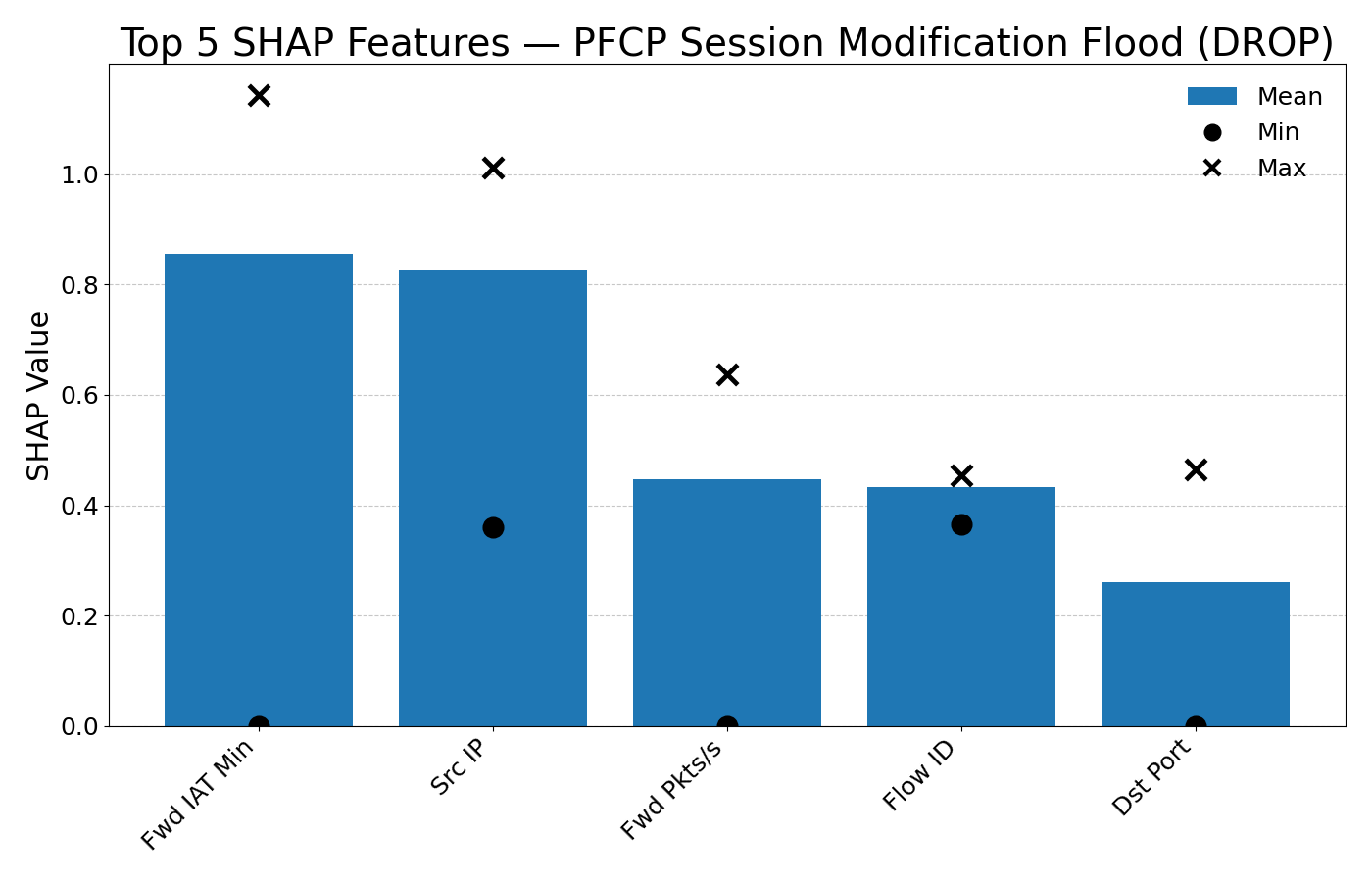}
    \caption{PFCP Session Modification Flood (DROP)}
    \label{fig:shap_3b}

\end{subfigure}

\caption{SHAP stability across different attacks: (a) UDPFlood, (b) SYNFlood, (c) Bidding down with TAUReject, (d) Authentication relay attack, (e) PFCP Session Deletion Flood, (f) PFCP Session Modification Flood (DROP).}
\label{fig:shap}
\end{figure}

Figure \ref{fig:shap} presents the five features with the highest mean SHAP values for given attack classes (x-axis). For each feature, the blue bar shows the mean SHAP value across all samples of that attack class (y-axis). The markers above and below the bar denote the maximum and minimum SHAP values observed for that feature, thereby indicating the stability of attributions. 

We adopt this visualization for conciseness and readability. While standard SHAP plots (e.g., force, waterfall, and beeswarm) are well-suited for analyzing individual predictions or global attribution distributions, they do not scale to class-level analysis and would require hundreds of figures to cover all the samples we studied for all the attack classes.

Considering the stability of SHAP in the three use cases, we see that SHAP values range from 0 to 8.28 for 5G-NIDD. For UDPFlood (Figure \ref{fig:shap_1a}), \textit{Seq} and \textit{Offset} hold the highest importance with moderate variability, while protocol-specific features (e.g., \textit{udp}) are less relevant. In DoS attacks (e.g. SYNFlood, Figure \ref{fig:shap_1b}), SHAP shows high variance, whereas scan-based attacks yield higher stability. 
 
For the MSA use case, SHAP values range from 0 to 2.91. Variability is greater than in 5G-NIDD: in both “Bidding down with TAUReject” (Figure \ref{fig:shap_2a}) and “Authentication relay attack” (Figure \ref{fig:shap_2b}), features often receive zero attribution for some instances but higher attribution for others, reflecting low stability. In these and the following charts, the features from this dataset are indicated as F1,..., F25 for readability. Their full feature names and descriptions appear in Table \ref{tab:features_MSA}.

SHAP values range from 0 to 1.17 for the PFCP use case. Explanations again show considerable variability: in the Session Deletion Flood (Figure \ref{fig:shap_3a}), \textit{Fwd IAT Min} is the most important, while \textit{Dst IP} and \textit{Fwd Pkts/s} fluctuate; in the Modification Flood (DROP) (Figure \ref{fig:shap_3b}), \textit{Fwd IAT Min} and \textit{Dst Port} {exhibit low stability.}

\begin{figure}[htbp]
    \centering
    
    \begin{subfigure}{0.48\textwidth}
        \includegraphics[width=\linewidth]{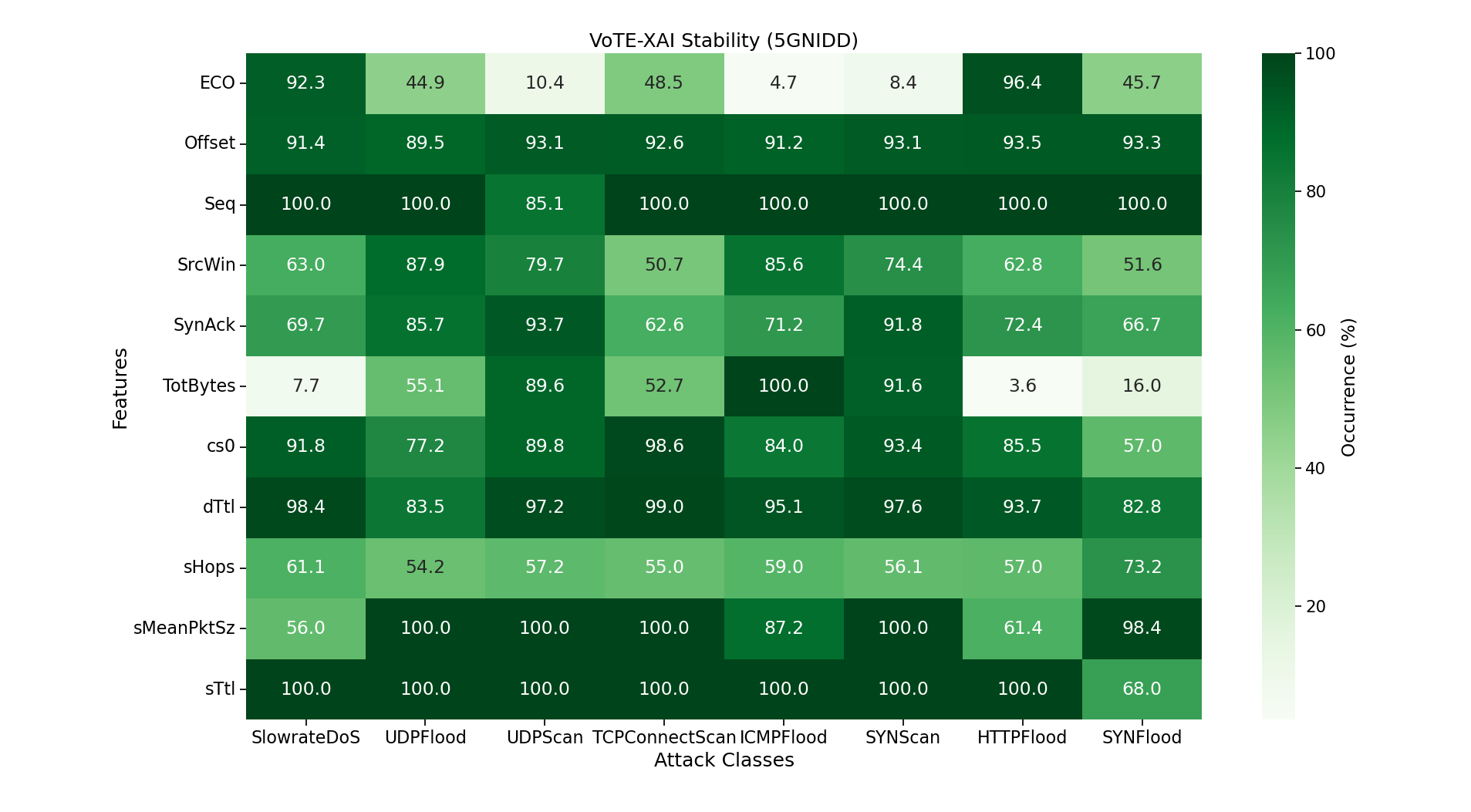}
        \caption{}
        \label{fig:vote_stab_1}
    \end{subfigure}
    \hfill
    \begin{subfigure}{0.48\textwidth}
        \includegraphics[width=\linewidth]{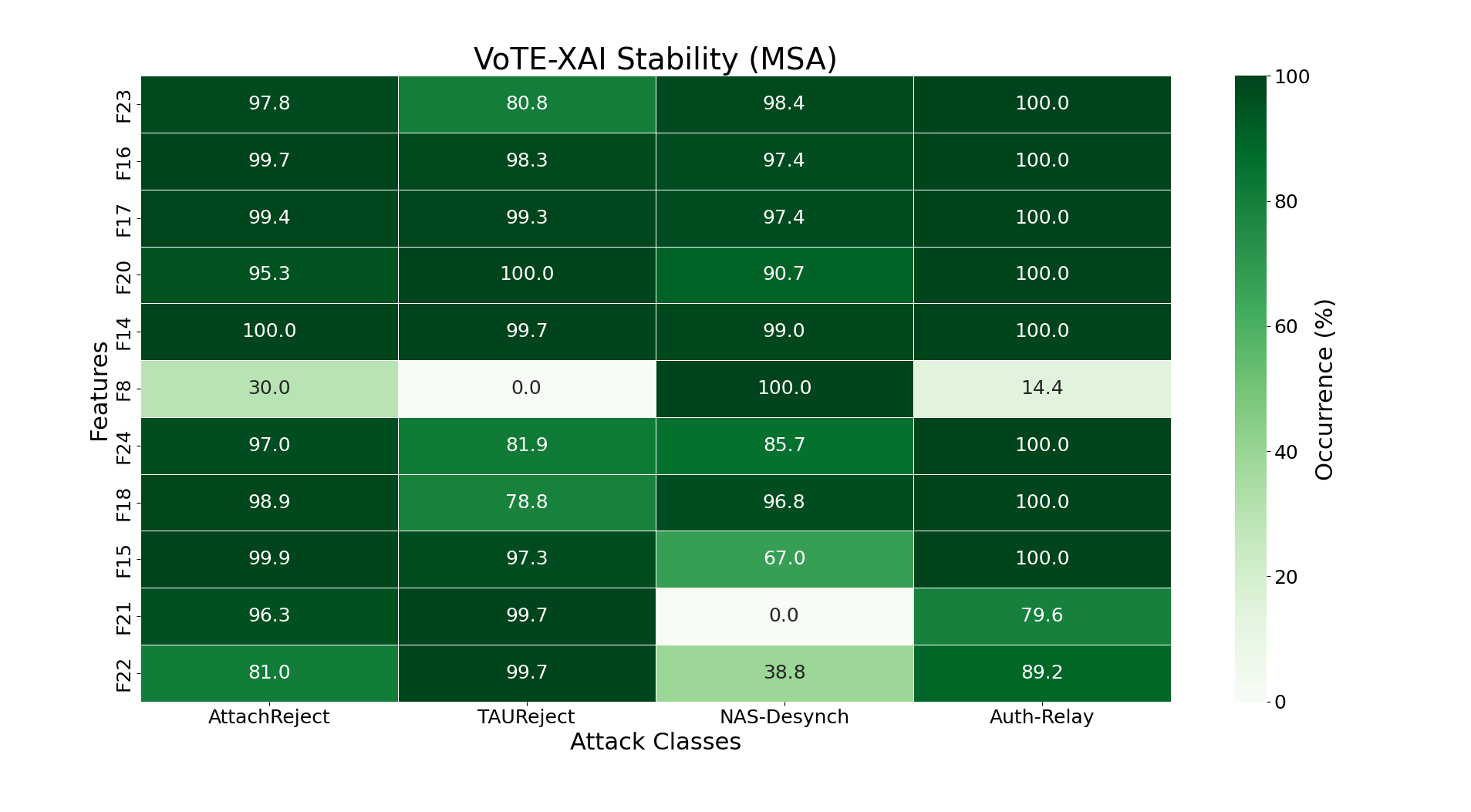}
        \caption{}
        \label{fig:vote_stab_2}
    \end{subfigure}
    \hfill
    \begin{subfigure}{0.5\textwidth}
        \includegraphics[width=\linewidth]{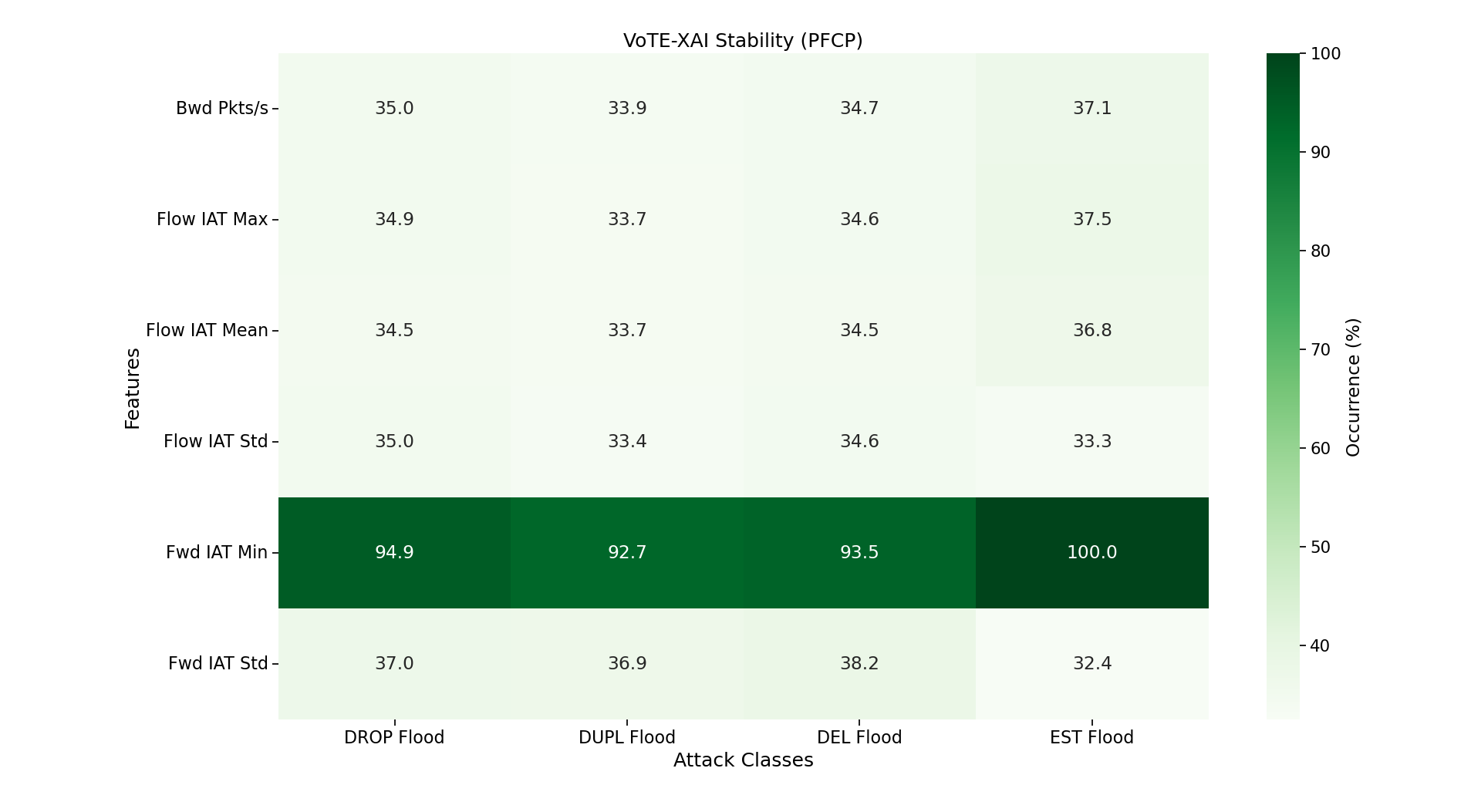}
        \caption{}
        \label{fig:vote_stab_3}
    \end{subfigure}

    \caption{VoTE-XAI stability for (a) 5G-NIDD, (b) MSA, and (c) PFCP datasets.}
    \label{fig:vote_stab}
\end{figure}

The heatmaps in Figure \ref{fig:vote_stab} visualize the stability of VoTE-XAI explanations across the three datasets by highlighting recurring feature patterns within each attack type. The x-axis refers to the attack classes, while the y-axis shows the features. The cells represent the occurrence (in \%) of the features across the explanations of all samples for the given attack type. Darker cells indicate features that appear with higher frequency.

\begin{table}[htbp]
\centering
\caption{Globally and attack-specific important features according to SHAP.}
\resizebox{\textwidth}{!}{%
\begin{tblr}{
  colspec = {|l|l|X[3]|l|},
  row{2} = {c},
  row{13} = {c},
  cell{2}{1} = {c=4}{},
  cell{3}{1} = {r=3}{},
  cell{6}{1} = {r=3}{},
  cell{9}{1} = {r=4}{},
  cell{13}{1} = {c=4}{},
  cell{14}{1} = {r=2}{},
  cell{16}{1} = {r=2}{},
  cell{18}{1} = {r=2}{},
  vlines,
  hline{1-3,6,9,13-14,16,18,20} = {-}{},
  hline{4-5,7-8,10-12,15,17,19} = {2-4}{},
}
\textbf{Use Case} & \textbf{Feature} & \textbf{Details} & \textbf{Importance} \\
\textbf{Global Importance} &  &  &  \\
5G-NIDD & Seq (Sequence Number) & Highest across all attack types & Mean SHAP: 3.1--4.3 \\
 & Offset (Fragmentation Offset) & High in UDPFlood, SYNScan, ICMPFlood & Mean: 3.24, 3.60, 3.26 \\
 & sTtl (Source TTL) & Relevant in UDPFlood, HTTPFlood, SYNFlood & Mean: 1.6--2.0 \\
MSA & NAS Security features & Across most attack types & Mean: 0.17--1.83 \\
 & NAS Identifiers \& E.212 & Half of attack types & Mean: 0.10--0.31 \\
 & GSM A / DTAP & Several DoS attacks & Mean: 0.63--1.31 \\
PFCP & Fwd IAT Min & All attack types & Mean: 0.71--0.98 \\
 & Dst IP & Frequent in top 5 & Mean: 0.55--0.69 \\
 & Src IP & Several classes & Mean: 0.51--0.83 \\
 & Fwd Pkts/s & Most classes & Mean: 0.40--0.45 \\
\textbf{Attack-Specific Importance} &  &  &  \\
5G-NIDD & sMeanPktSz & SYNFlood & Mean: 3.80; Min: 1.05; Max: 8.28 \\
 & TotBytes & ICMPFlood & Mean: 5.99 (constant) \\
MSA & NAS Security (F23) & NAS Counter Desynchronization & Mean: 0.20; Max: 0.27 \\
 & NAS Identifiers \& E.212 (F25) & Incarceration with rrcReestablishReject & Mean: 0.09; Max: 0.24 \\
PFCP & Pkt Len Std & Session Establishment Flood & Mean: 0.586 (constant) \\
 & Dst Port & Session Modification Flood (DROP) & Mean: 0.26; Min: 0.00; Max: 0.46 \\
\end{tblr}
}
\label{tab:SHAP_stab}
\end{table}

\begin{table}[htbp]
\centering
\caption{Globally and attack-specific important features according to VoTE-XAI}
\label{tab:Vote_stab}
\resizebox{\textwidth}{!}{%
\begin{tblr}{
  row{2} = {c},
  row{11} = {c},
  cell{2}{1} = {c=4}{},
  cell{3}{1} = {r=3}{r},
  cell{6}{1} = {r=2}{r},
  cell{8}{1} = {r=3}{r},
  cell{11}{1} = {c=4}{},
  cell{12}{1} = {r=4}{r},
  cell{16}{1} = {r=2}{r},
  cell{18}{1} = {r},
  vlines,
  hline{1-3,6,8,11-12,16,18-19} = {-}{},
  hline{4-5,7,9-10,13-15,17} = {2-4}{},
}
\textbf{Use Case}                   & \textbf{Feature}                                                 & \textbf{Details}                                                               & \textbf{Importance / Occurrence}       \\
\textbf{Global Importance}          &                                                                  &                                                                                &                                        \\
5G-NIDD                             & Seq                                                              & All minimal explanations (except UDPScan)                                      & 100\% (85\% in UDPScan)                \\
                                    & sTtl                                                             & All attack types (except SYNFlood)                                             & 100\% (68\% in SYNFlood)               \\
                                    & Offset                                                           & Top feature across attack types                                                & Near-perfect                           \\
MSA                                 & {NAS Security, \\NAS EPS EMM, \\NAS Message Internals  Encoding} & Nearly all attack classes                                                      & 100\%                                  \\
                                    & NAS Identifiers  E.212                                           & {Authentication Relay, \\Bidding Down with AttachReject, \\Handover Hijacking} & 99–100\%                               \\
PFCP                                & Fwd IAT Min                                                      & Almost every attack class                                                      & 93–100\%                               \\
                                    & Flow IAT Max, Flow IAT Std                                       & Multiple attack types                                                          & 33–38\%                                \\
                                    & Bwd/Pkts per Second                                              & All attacks                                                                    & 34–37\%                                \\
\textbf{Attack-Specific Importance} &                                                                  &                                                                                &                                        \\
5G-NIDD                             & ECO (Ping request)                                               & {SlowrateDoS, HTTPFlood; low in scans and \\ICMPFlood}             & {92.3\%, 96.4\%; \\low: 4.6-10.4\%} \\
                                    & SynAck (SYN-ACK Time in Handshake)                               & SYNScan, UDPFlood, ICMPFlood,                                                              & 72.4\%-93.7\%                       \\
                                    & sMeanPktSz (Mean Packet Size)                                    & SYNScan, UDPScan, TCPConnectScan, UDPFlood                                     & 100\%                                  \\
                                    & cs0 (Class of Service)                                           & Scan attacks, SlowrateDoS                                                      & {89\% in UDPScan\\91\% in SlowrateDoS} \\
MSA                                 & F8                                                & Only NAS Desynch                                               & Near perfect frequency                 \\
                                    & F21                                          & Bidding Down via TAUReject and AttachReject                                                    & Consistently high                      \\
PFCP                                & –                                                                & No attack specific features identified                                         & –                                    
\end{tblr}
}
\end{table}

In 5G-NIDD, VoTE-XAI highlights a consistent core of features per attack class for all minimal explanations per sample (Figure \ref{fig:vote_stab_1}). Some features consistently appear in over 80\% of the explanations across multiple attack types, indicating their high relevance in distinguishing malicious activity (e.g. \textit{Seq}, \textit{sTtl}, \textit{Offset}). Certain features appear consistently in specific attack types (e.g. \textit{ECO} has near-perfect frequency only for SlowrateDoS and HTTPFlood). 

For the MSA dataset, with the 478 dimensional model, the numbers on the charts are based on the calculation of all minimal explanations found by VoTE-XAI with a 1 hour timeout. As shown in Figure \ref{fig:vote_stab_2}, VoTE-XAI maintains high intra-class consistency, distinguishing globally relevant features (indicative of protocol-level malicious behavior) from those specific to individual attack types. For brevity, the heatmap for this use case shows a subset of four representative attacks out of the full set of ten. 

In the PFCP use case, the one with the smallest number of features, all minimal explanations were generated. VoTE-XAI explanations show a high degree of stability, calculated based on all minimal explanations per sample: all four attack types share a near-identical core set of features, with negligible differences in terms of occurrence for DROP and DUPL attacks (Figure \ref{fig:vote_stab_3}). This consistency is expected, as the underlying traffic characteristics of PFCP floods are similar across variants.

\subsection{Divergence Among the Methods}
\label{sub:divergence}
The results in the last subsection reveal significant divergence between SHAP and VoTE-XAI. Several features that appear in over 80\% of VoTE-XAI explanations have low or zero SHAP attribution, indicating a fundamental misalignment between the methods. 

Table \ref{tab:div} details the divergence between SHAP attribution and VoTE-XAI occurrences.

\begin{table}[htbp]
\centering
\caption{Summary of divergence between VoTE-XAI and SHAP feature importance across use cases.}
\label{tab:div}
\resizebox{\textwidth}{!}{%
\begin{tblr}{
  row{1} = {c},
  cell{2}{1} = {r=6}{r},
  cell{8}{1} = {r=2}{r},
  cell{10}{1} = {r=4}{r},
  vlines,
  hline{1-2,8,10,14} = {-}{},
  hline{3-7,9,11-13} = {2-4}{},
}
\textbf{Use Case} & \textbf{Feature(s)}                                      & \textbf{VoTE-XAI Observation}                                    & \textbf{SHAP Observation}                                                 \\
5G-NIDD           & dTtL~                                                    & {Critical in multiple attacks \\(82.8–99\% occurrence)}          & {Mean importance = 0 for most;\\~0.64 in SlowrateDoS, \\0.48 in SYNFlood} \\
                  & ECO~                                                     & {92.3\% in SlowrateDoS; 96\% \\in HTTPFlood}                     & Mean importance $\approx$ 0.24                                            \\
                  & TotBytes                                                 & {$\approx$92\% in SYNScan\\minimal explanations}                 & Mean importance $\approx$ 0                                               \\
                  & cs0~                                                     & {Appears in almost all minimal \\explanations}                   & {Near 0 importance \\across all attacks}                                  \\
                  & SynAck~                                                  & {85.7–93.7\% in SYNFlood, \\TCPConnectScan, UDPScan}             & Near 0 mean value                                                         \\
                  & General trend                                            & {All top-ranked SHAP features \\appear in VoTE-XAI explanations} & {VoTE-XAI-only features \\often missed}                                   \\
MSA               & {GSM A / DTAP, \\NAS Security, \\NAS Identifiers  E.212} & {100\% inclusion in \\Authentication Relay}                      & Mean importance~$<$ 0.07                                                  \\
                  & {NAS EPS EMM, \\NAS Message Internals  Encoding}         & {100\% in Bidding Down \\with AttachReject}                      & {Mean importance $\approx$ 0.14,\\0.11}                                   \\
PFCP              & Fwd IAT Min                                              & 92.7\% occurrence in DUPL flood                                  & {Mean SHAP: 0.80~\\(agreement)}                                           \\
                  & Bwd Pkts/s                                               & {33.9\% in DUPL; frequent across \\floods}                       & Mean SHAP: 0.0023                                                         \\
                  & Flow IAT Max                                             & {33.7\% in DUPL; 37.5\% in \\Session Establishment Flood}        & Mean SHAP: 0.0009                                                         \\
                  & Flow IAT Std                                             & $>$33\% occurrence across floods                                 & Near-zero attribution                                                     
\end{tblr}
}
\end{table}

\begin{figure}[htbp]
\centering

\begin{subfigure}[t]{0.45\textwidth}
    \centering
    \includegraphics[width=\linewidth]{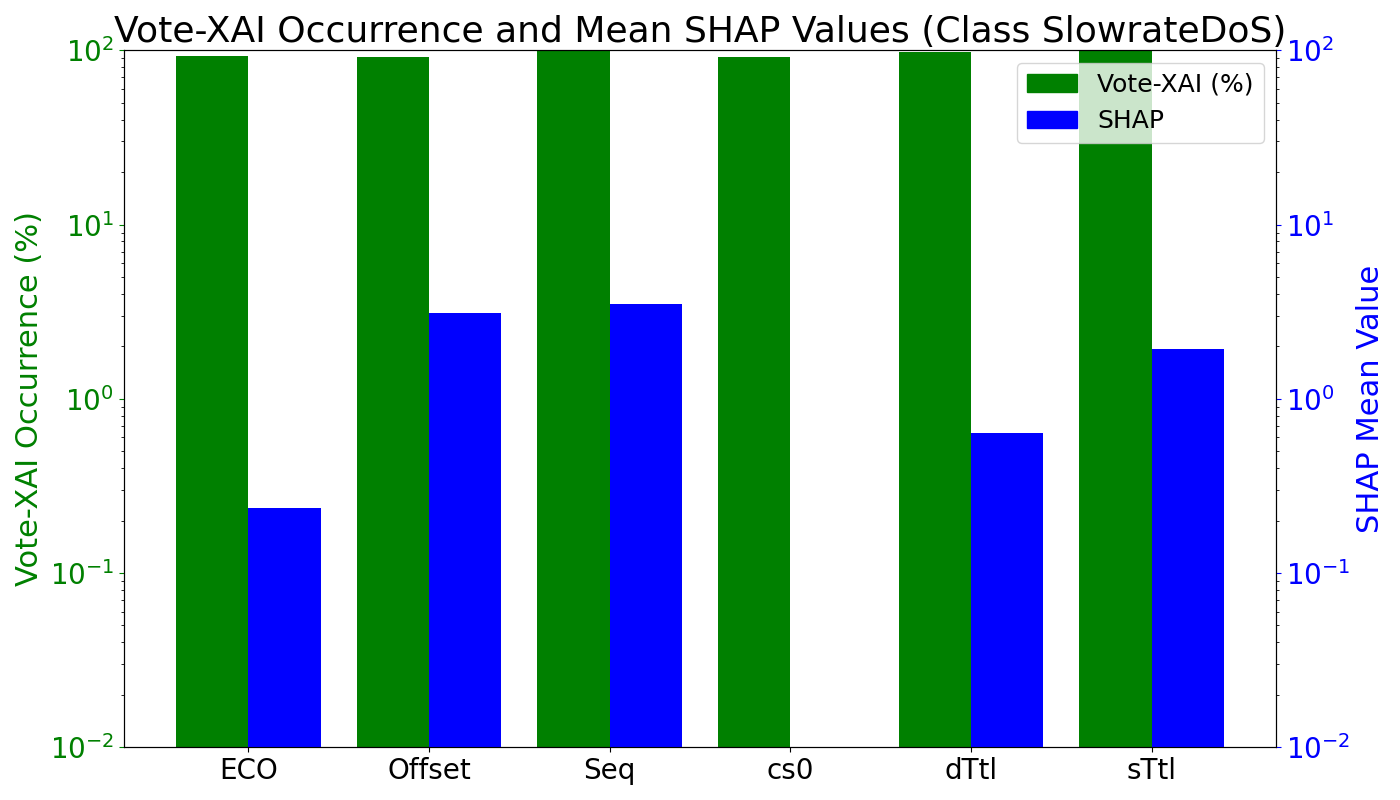}
    \caption{SlowrateDoS}
    \label{fig:vote_1a}
\end{subfigure}
\hfill
\begin{subfigure}[t]{0.45\textwidth}
    \centering
    \includegraphics[width=\linewidth]{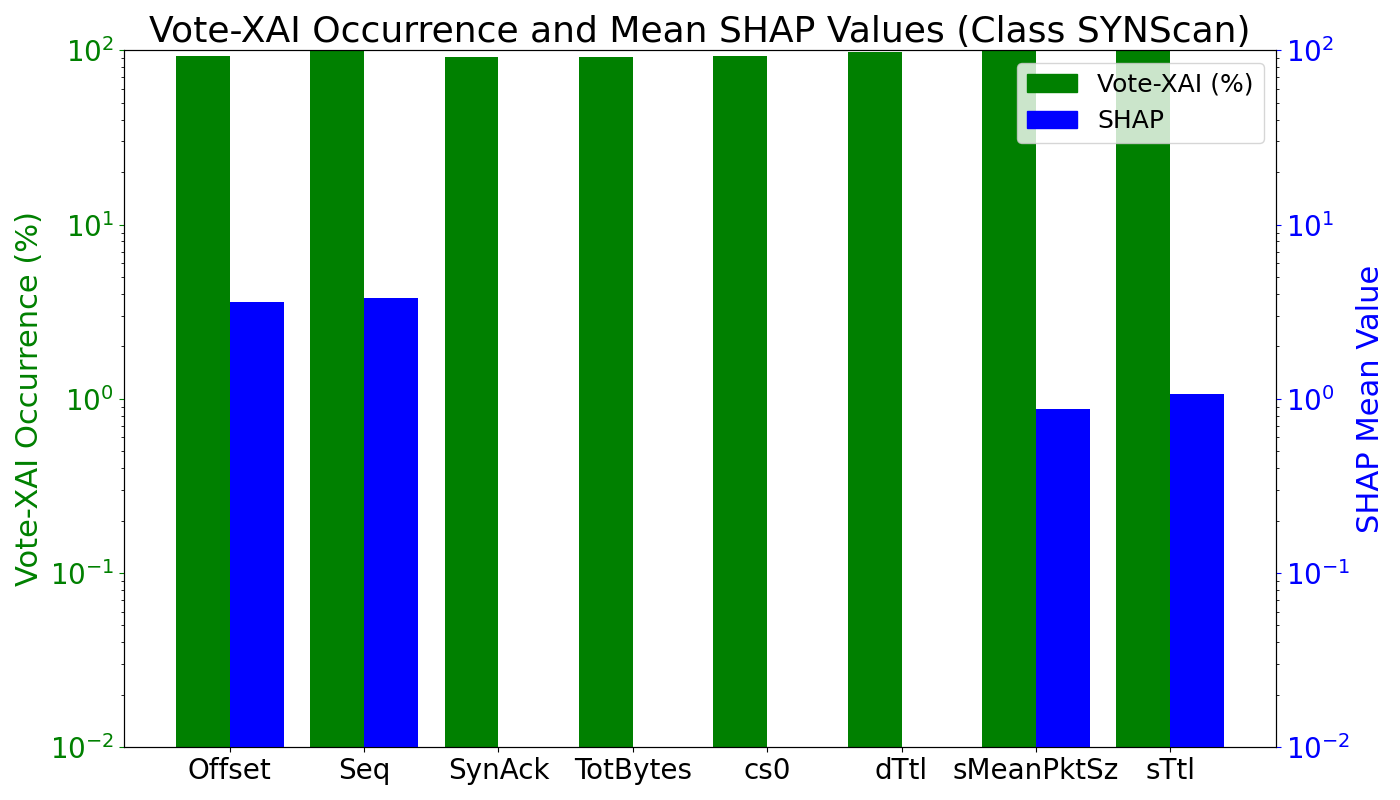}
    \caption{SYNScan}
    \label{fig:vote_1b}
\end{subfigure}

\vskip\baselineskip

\begin{subfigure}[t]{0.48\textwidth}
    \centering
    \includegraphics[width=\linewidth]{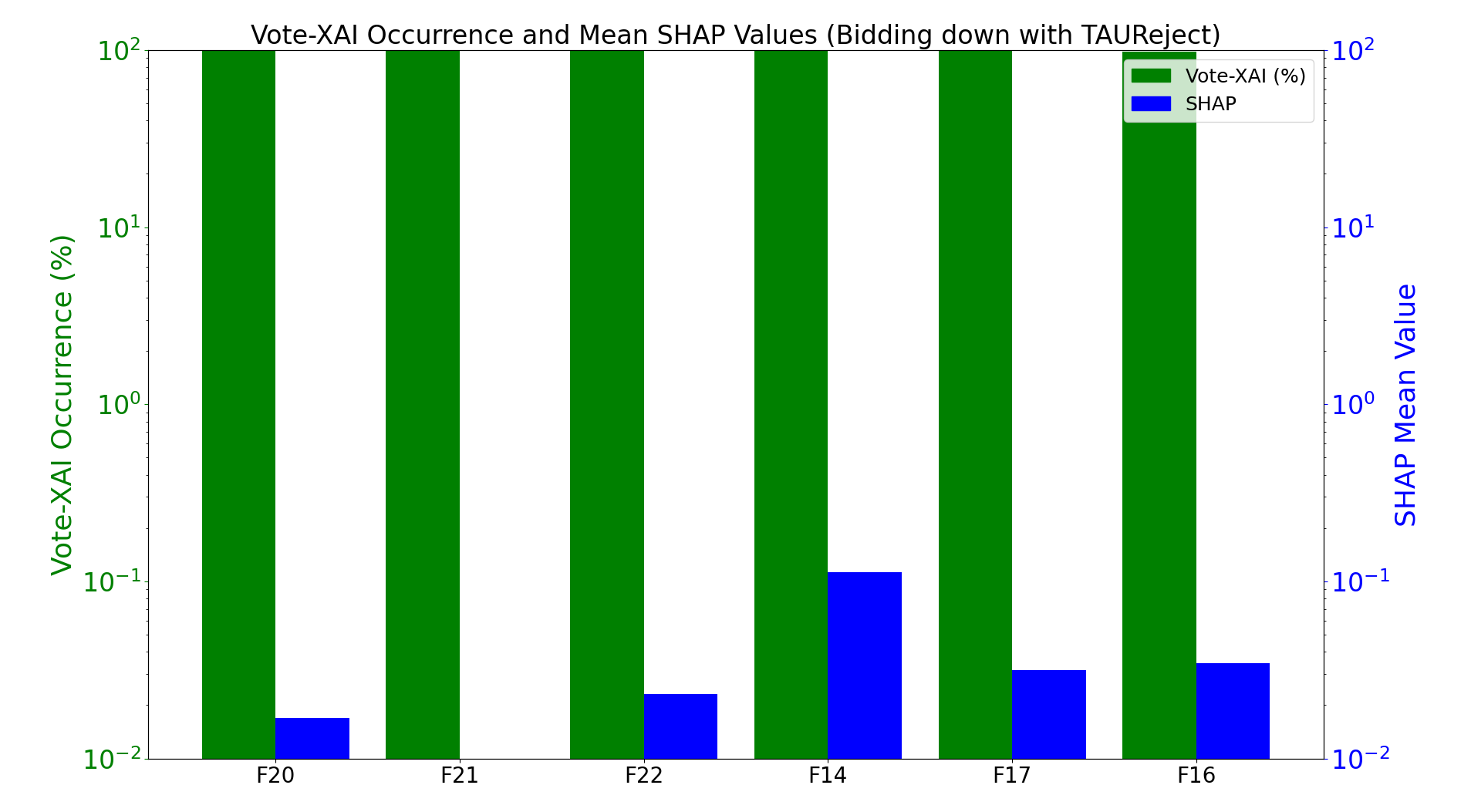}
    \caption{Bidding down with TAUReject}
    \label{fig:vote_2a}
\end{subfigure}
\hfill
\begin{subfigure}[t]{0.48\textwidth}
    \centering
    \includegraphics[width=\linewidth]{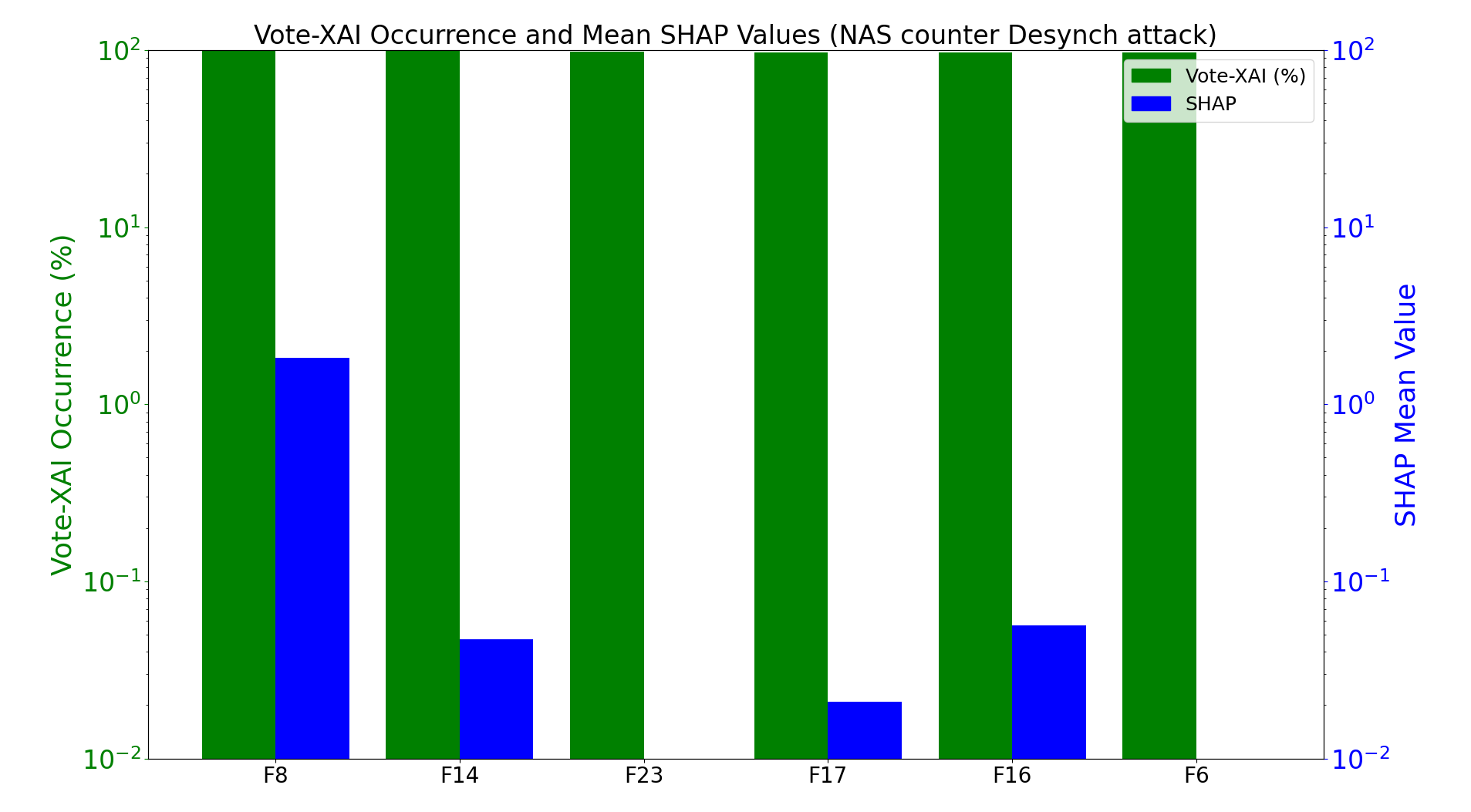}
    \caption{NAS Counter Desynch attack}
    \label{fig:vote_2b}
\end{subfigure}

\vskip\baselineskip

\begin{subfigure}[t]{0.48\textwidth}
    \centering
    \includegraphics[width=\linewidth]{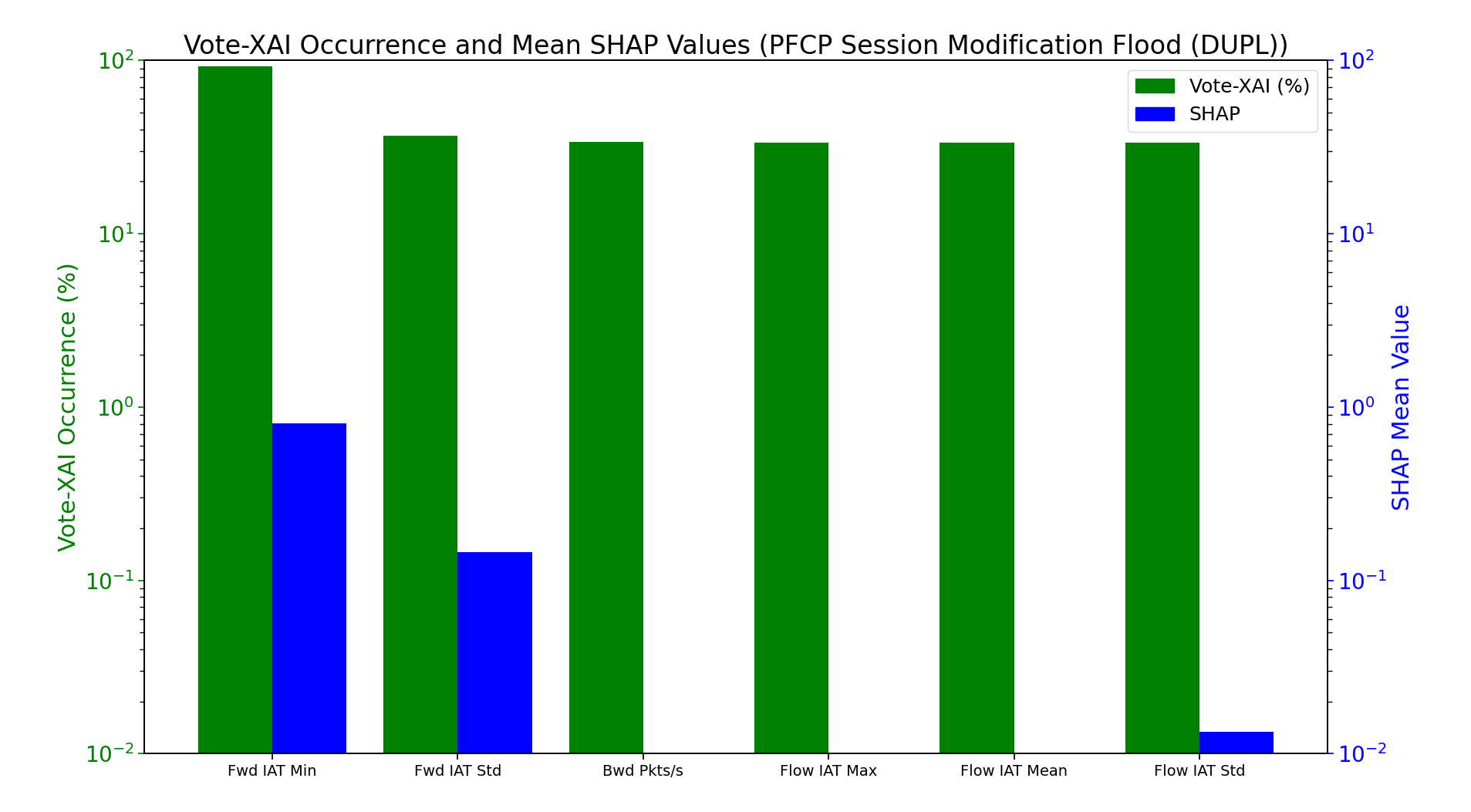}
    \caption{PFCP Session Modification Flood (DUPL)}
    \label{fig:vote_3a}
\end{subfigure}
\hfill
\begin{subfigure}[t]{0.48\textwidth}
    \centering
    \includegraphics[width=\linewidth]{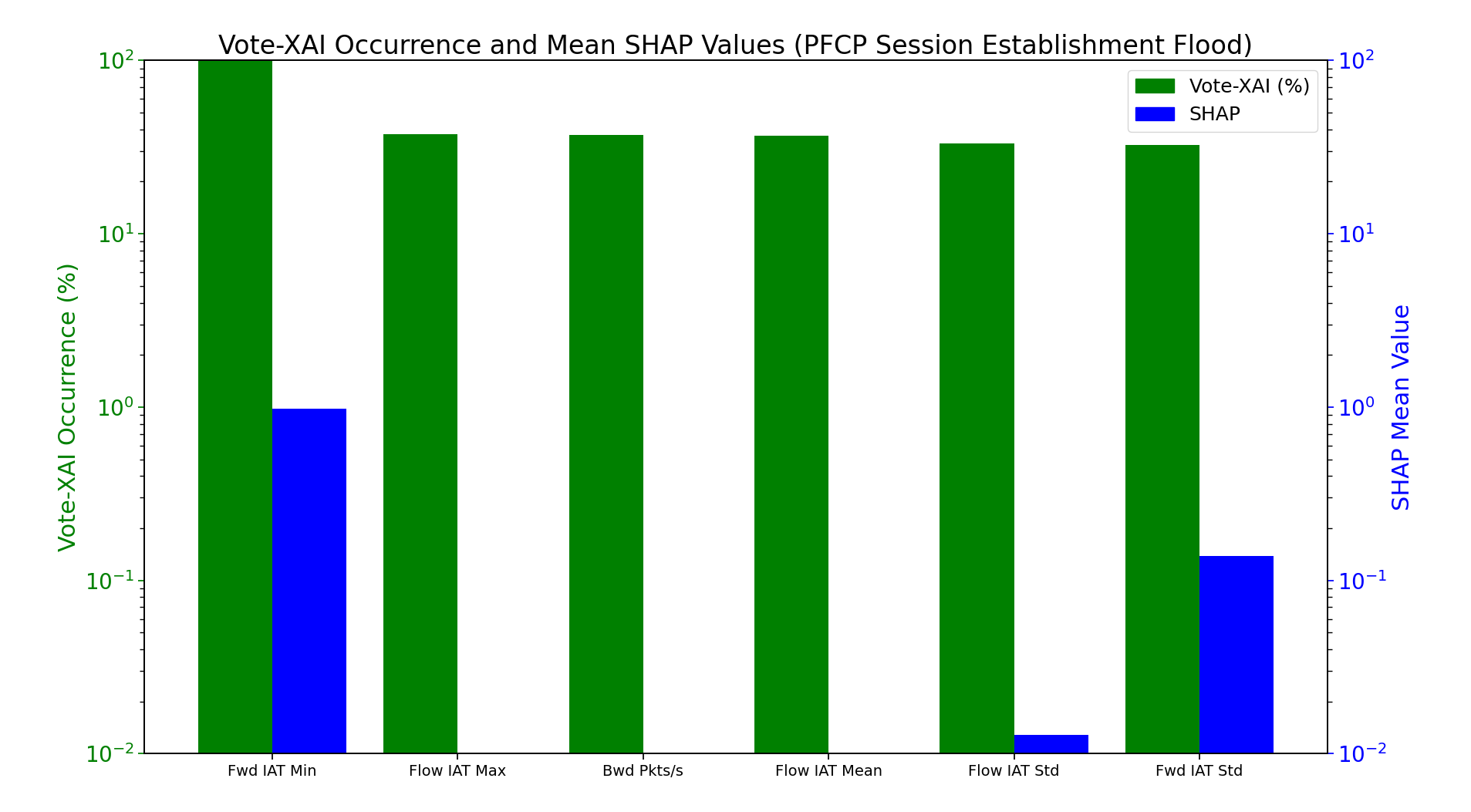}
    \caption{PFCP Session Establishment DoS}
    \label{fig:vote_3b}
\end{subfigure}

\caption{Divergence between VoTE-XAI feature importance (green bars) and SHAP attribution (blue bars) across six attack types. The y-axes are in logarithmic scale.}
\label{fig:vote}
\end{figure}

Figure \ref{fig:vote} highlights representative cases where VoTE-XAI and SHAP exhibit significant discrepancies in feature attribution. 
In the charts, the green (left) y-axis represents VoTE-XAI occurrence (in \%), while the blue one (right) represents the mean SHAP values. The x-axis shows the features most frequently selected by VoTE-XAI (occurrence $>$80\%; for the PFCP dataset, a lower threshold of 30\% was used for visualization).
It is worth mentioning that a similar trend was observed in each attack class, with SHAP systematically assigning near-zero scores to at least one of the top VoTE-XAI features. 

\textbf{5G-NIDD.} In SlowrateDoS (Figure \ref{fig:vote_1a}), features such as \textit{dTtL} (occurrence 98.4\%) and \textit{ECO} (92.3\%) are consistently selected by VoTE-XAI but underestimated by SHAP (0 and 0.24, respectively), despite being critical indicators of slow-rate attacks. In SYNScan (Figure \ref{fig:vote_1b}), VoTE-XAI highlights \textit{dTtL} (97\%) and \textit{TotBytes} (91.6\%), while SHAP again assigns them zero importance. 

\textbf{MSA.} Similar discrepancies appear in signaling-centric and identity manipulation attacks. For instance, in ``Bidding down with TAUReject'' (Figure \ref{fig:vote_2a}) and the ``NAS counter Desynch attack'' (Figure \ref{fig:vote_2b}), VoTE-XAI consistently selects protocol-level fields (e.g., NAS identifiers, encoding details) with high frequency, while SHAP attributes them near-zero importance. Comparable patterns are observed in other MSA attack classes.

\textbf{PFCP.} Disagreements are also evident in this use case. For PFCP Session Modification Flood (DUPL) (Figure \ref{fig:vote_3a}), both methods align on \textit{Fwd IAT Min} (92.7\%, SHAP: 0.81), but diverge on others: VoTE-XAI highlights \textit{Bwd Pkts/s} (33.9\%, SHAP: 0.0023) and \textit{Flow IAT Max} (33.7\%, SHAP: 0.0009). In Session Establishment DoS (Figure \ref{fig:vote_3b}), VoTE-XAI consistently emphasizes \textit{Flow IAT Max} and \textit{Bwd Pkts/s}, both overlooked by SHAP. Similar trends hold across other PFCP floods.

Several features consistently highlighted by VoTE-XAI but largely ignored by SHAP correspond to detection indicators described in recent literature \cite{hirsi2025comprehensive}. For example, in SYN-based flooding and scanning, VoTE-XAI highlighted SYN–ACK handshake timing, an indicator recognized in detecting reconnaissance and handshake abuse. Instead, SHAP emphasizes variables with statistical correlation but weaker causal interpretability (e.g., generic byte counts).

These results answer RQ1 by systematically identifying the divergent behaviors of the two approaches on the same model across three use cases. Section \ref{sec:discussion} provides a summarizing discussion.

\subsection{Efficiency Evaluation}
\label{sub:time}
\begin{figure}[htbp]
    \centering

    \begin{subfigure}[t]{0.5\textwidth}
        \centering
        \includegraphics[width=\linewidth]{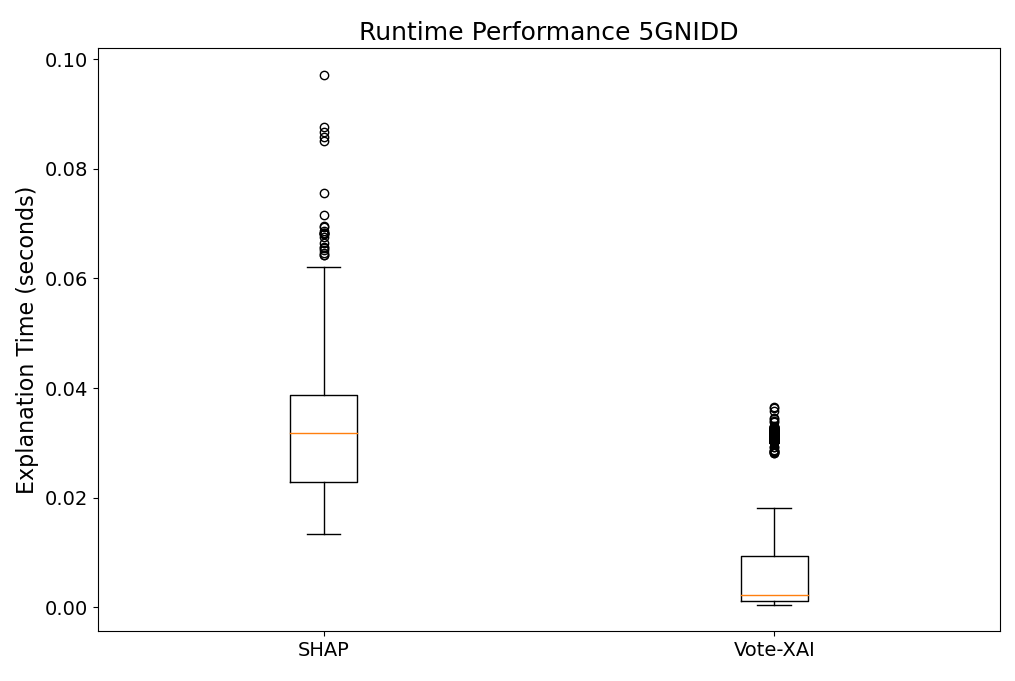}
        \caption{}
        \label{fig:time_1}
    \end{subfigure}

    \vskip\baselineskip
    \begin{subfigure}[t]{0.48\textwidth}
        \centering
        \includegraphics[width=\linewidth]{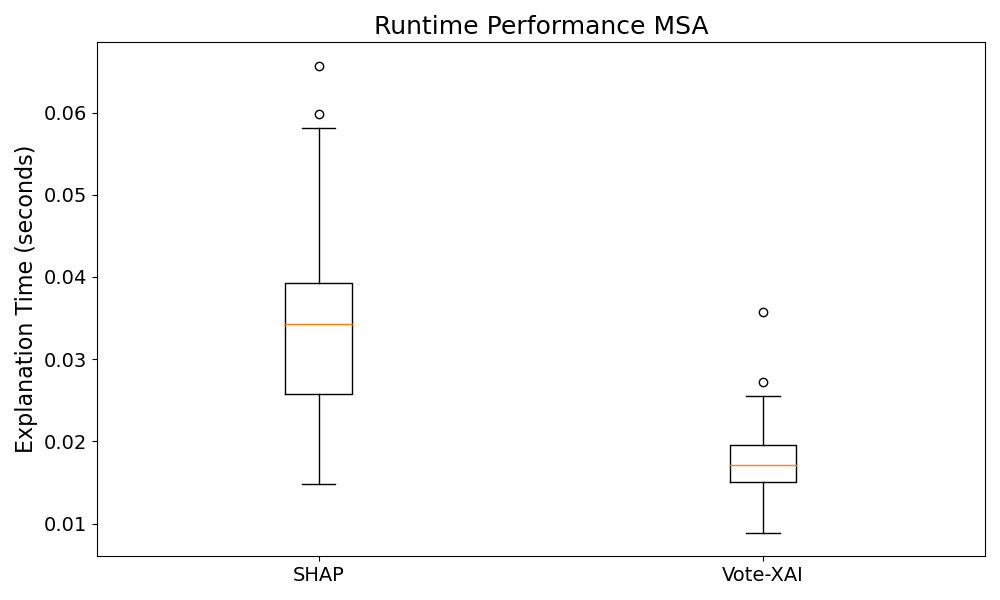}
        \caption{}
        \label{fig:time_2}
    \end{subfigure}
    \hfill
    \begin{subfigure}[t]{0.48\textwidth}
        \centering
        \includegraphics[width=\linewidth]{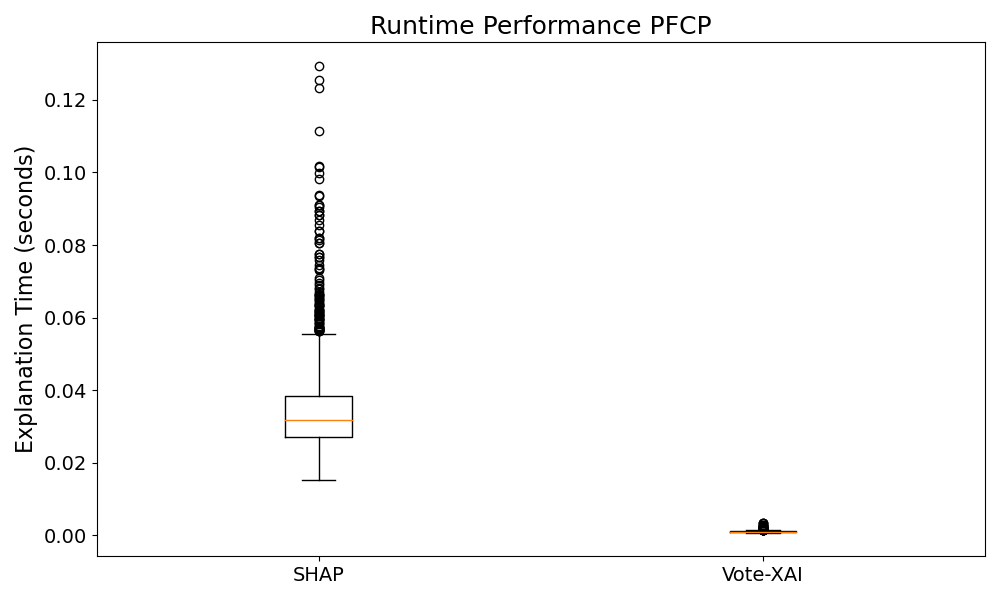}
        \caption{}
        \label{fig:time_3}
    \end{subfigure}

    \caption{
    Efficiency Evaluation for SHAP and VoTE-XAI (single minimal explanation) for the 5G-NIDD (a), MSA (b), and PFCP (c) datasets.}
    \label{fig:time}
\end{figure}

This section evaluates the 
efficiency of SHAP and VoTE-XAI explanations.
The results, shown in Figure \ref{fig:time_1}, \ref{fig:time_2}, and \ref{fig:time_3}, compare the SHAP runtime to the time required to compute one minimal explanation per sample using VoTE-XAI.

For 5G-NIDD, the mean runtime for SHAP is  0.032 seconds, whereas VoTE-XAI demonstrates a significantly lower mean runtime (0.002 seconds). Additionally, the outliers for VoTE-XAI are fewer and lower than the SHAP outliers.

A similar trend is observed in the MSA use case, with a mean runtime of 0.034 seconds for SHAP, and  0.0017 seconds for VoTE-XAI, and in the PFCP one (0.03 seconds for SHAP, and 0.001 seconds for VoTE-XAI). 

However, computing \emph{all} minimal explanations for a single alert is significantly more time-consuming, which we do not show by charts here but only summarize, since it can only be done by VoTE-XAI. VoTE-XAI requires between 0.5 and 9 seconds per sample for the 5G-NIDD dataset, and between 0.6 and 1.8 seconds for the PFCP dataset. During this time, hundreds of minimal and provably correct explanations are computed per sample. Note that Figure \ref{fig:vote} showed the convergence of the features appearing in those different minimal explanations to a great extent.

For the MSA dataset, given the larger feature space and thereby a much larger search space, the full computation of all minimal explanations for a single attack class required approximately 17 hours on average. Due to this high computational demand, a timeout of one hour was set when computing all minimal for the stability and divergence analyses presented in previous sections. We will discuss this further in Section \ref{sec:discussion}.

\subsection{False Positive Analysis}
\label{sub:fp}
To complement the main TP-focused evaluation, we performed a detailed review of all FPs cases in the 5G-NIDD dataset, which is the one with the highest number of FPs. Yet, as shown in Table \ref{tab:performance}, the model produced only 37 FPs (benign flows misclassified as malicious).
For benign flows that were flagged, both methods surface traffic-shape and basic header cues. SHAP emphasizes byte/size and TTL indicators together with simple protocol markers (\texttt{TotBytes}, \texttt{sTtl}, \texttt{sMeanPktSz}, \texttt{INT}, \texttt{Offset}). VoTE-XAI frequently includes \texttt{Seq}, but does not bring in path/handshake structure (e.g., \texttt{dTtl}, \texttt{SynAck}, \texttt{dHops}). This suggests that the FPs could be benign bursts that mimic DoS-like shape, without the footprint seen in actual attacks.
In light of the experiments, a consistent pattern emerges: (i) for TPs, traffic-shape features co-occur with sequence/fragmentation and path/handshake cues in both methods (for the UDPFLood class); (ii) FPs lack this structural footprint. This information may be useful for a domain expert to de-prioritize alerts that most likely are FPs and appear at run-time.

\section{Discussion}
\label{sec:discussion}
The above experimental results provide some insights regarding trade-offs between statistical and logic-based feature attribution methods. 
While SHAP is grounded in a model-agnostic framework, in our case, we used its TreeExplainer module, optimized for tree ensembles. This optimization improves computational efficiency and ensures fair allocation of feature contributions, yet the resulting explanations still tend to be less sparse and less stable in this concrete security setting, as we discuss below.

\subsection{Sparsity and Efficiency}
We saw that VoTE-XAI provides {sparser} feature attributions, which is significant for mitigation in systems with large feature sets. Regarding efficiency, VoTE-XAI is significantly faster than SHAP, making it suitable for real-time security monitoring. 

When asked to calculate \emph{all} minimal explanations for a given alert, the runtime of VoTE-XAI can increase significantly, as shown in the case of the MSA dataset. 
This highlights a potential trade-off: VoTE-XAI is capable of providing a more comprehensive analysis if additional computation time is available, but already with one minimal explanation directs to logically valid explanations, given a model.

In the context of intrusion detection, where real-time response is critical, this trade-off raises an important question: is the added computational cost justified by the additional explanations? Given that DoS attacks can cause significant damage within seconds or minutes \cite{de2023distributed}, further delay might be challenging for immediate mitigation. 

On the other hand, for other attacks, and because real network outages can take hours or even days to get sorted out, it may not be an issue to spend the time required to have a full picture (in our use case 1, an additional 9 seconds). While this delay may or may not work in a live detection scenario, it appears plausible in a forensic context.

\subsection{Stability and Correctness}
Beyond sparsity and efficiency, a crucial aspect to consider is the reliability of explanations. Reliability rests on stability, meaning that we expect similar explanations over instances of the same attack class. 

We saw that VoTE-XAI provides more stable feature attributions, whereas SHAP exhibits lower stability and, in some cases, fails to highlight security-critical features (Section \ref{sub:stab}). 

Notably, even when VoTE-XAI computes a single (minimal) explanation, the high stability of explanations among attack classes (as seen in Section \ref{sub:stab}) provides some confidence that the security team can act on trustworthy information.

Even with stable explanations, a further concern is correctness.
In the 5G intrusion context, correctness refers to semantic validity: do the explanations align with known attack mechanics? As shown in Section \ref{sub:divergence}, this is the case for multiple features deemed important by VoTE-XAI and overlooked by SHAP.

Consequences of inaccurate explanations are missed opportunities to pinpoint the cause of an attack or a waste of resources looking for irrelevant mitigation actions. This paper identifies the potential for some such explanations when using SHAP.

\subsection{Threat to Validity}

A major challenge in 5G security research is the lack of large-scale, diverse, and up-to-date datasets specifically curated for intrusion detection.
While the 5G-NIDD dataset is one of the most comprehensive available, it has inherent limitations in diversity and volume, particularly regarding protocol-specific threats. While relevant, the focus on Scan and DoS attacks represents a limited subset of potential 5G security threats.

To partially address this limitation, we incorporate the MSA and PFCP datasets, which focus on more sophisticated and protocol-aware control plane attacks, but are much smaller in terms of sample numbers. 

As highlighted by Arp et al. \cite{arp2024pitfalls}, ML in security research is prone to several pitfalls that can undermine the generalizability and reliability of results. Specifically, our use of pre-collected datasets exposes us to sampling bias and imbalance. However, since we focus only on explanations, we believe that the lessons learned can be transferred to new collection setups. 

Another threat relates to label inaccuracies arising due to reliance on heuristics or imperfect labeling mechanisms, but the impact of that is limited. Our method was demonstrated on true positive alerts, no matter how they are identified.

It is also important to acknowledge that the metrics selected for our evaluation focus on a subset of the explainability spectrum. Incorporating additional metrics may reveal complementary aspects of the two approaches.

This work suggests the need for realistic testbeds to study different attack and defense patterns, and the release of datasets to assess the usefulness and novelty of new AI-related tools in the security toolbox. In particular, a testbed would allow us to address part of the aforementioned limitations by enabling the creation of large and diverse datasets. This would enable studying attacks not addressed in this work and also make statistical analysis of feature attribution outcomes possible. Additionally, it would be a modular environment where components can be updated over time, allowing evaluation to remain relevant along the evolution of network architectures and compliant with new 3GPP updates.

\subsection{Towards Usability Studies}
Several areas warrant exploration to further enhance the applicability of XAI in 5G security.
Future research could focus on qualitative analysis of the methods through usability studies.

To simulate domain expert interpretation in 5G security, we have conducted experiments \cite{uccello2026mission} using a Large Language Model (LLM), Open-Mistral-7B\footnote{\url{https://docs.mistral.ai/}} in a different study. The LLM was prompted to extract mitigation assistance from logic-based feature attribution. Then, the mitigation actions were validated using expert-curated MITRE Five-G Hierarchy of Threats
(FiGHT) \cite{FiGHT2025}, with encouraging results.
While LLM outputs cannot replace expert judgment, {this preliminary research} shows that VoTE-XAI might have potential in highlighting actionable threat indicators. Our ongoing work studies the 5G-specific expert knowledge embedded in LLM for further study.

\section{Conclusion and Future Directions}
\label{sec:conclusion}

This study evaluated statistical (SHAP) and logic-based (VoTE-XAI) feature attribution methods for 5G security monitoring. 
The evaluation focused on metrics reflecting operational requirements, namely sparsity, stability, and efficiency.

Our findings show that VoTE-XAI consistently yields sparser attributions than SHAP and exhibits higher stability across multiple attack classes. While sparsity is a critical requirement in high-dimensional settings, we relate stability to semantic correctness, i.e., the consistent identification of features that meaningfully characterize a given attack type. In this context, we identified substantial disagreement between the two methods in terms of attack-specific features. In our experiments, VoTE-XAI attributions tended to emphasize attack-specific indicators, whereas SHAP attributions more frequently highlighted generic network features.
 
VoTE-XAI attributions (while logically correct relative to the detection model) are also shown to be computationally more efficient when generating a single minimal explanation, making them more suitable for real-time intrusion detection where sub-second calculations are needed. 

We have also assessed efficiency under a different lens by computing \emph{all} minimal explanations. This task incurs additional time, but it will most likely provide additional benefits in a forensic context. Additionally, instead of all minimal explanations, it would still be possible to compute a limited number of explanations based on time constraints or operational urgency. This hybrid strategy would allow analysts to balance completeness and latency, adapting the explanation granularity to the context.

Incorporating domain expert security knowledge into both methods could further improve the efficiency of explanations.
This is already foreseen in the computation engine of VoTE-XAI, which incorporates a cost function for more actionable features (but we ignored by setting all costs to 1 in these experiments). For SHAP values, threshold levels (above zero) would have to be experimentally determined.

Note that logically correct in this context means correct with respect to the ML model's perception of what is an attack and what is benign. However, our evaluations were done on the basis of known attacks in the data, correctly labeled by the model as a TP.
It is also worth to remark that the study used XGBoost as a baseline detector, and other model families might reveal different attribution patterns. In any case, given that ML models are never perfect, we can only come up to the level of accuracy of a model in terms of the relevance of attributions. In other words, given a true positive outcome of a detection, the attributed features by VoTE-XAI are as accurate as we can get. 

For a false positive outcome, further analyst investigation is typically required. This is not trivial, since such cases by definition reside on the border between attacks and benign data. Explainability can play an important role in this context by revealing which features or conditions drive borderline decisions. Initial approaches in the attack evasion context have been explored elsewhere \cite{colaco2024fast}, whereby a subset of flows can be prioritized lower by the operator, given the calculation of abstract regions before run-time deployment.

A critical next step is applying the proposed methodology within a realistic testbed environment, where we can control, inject, and monitor traffic in real-time while dynamically analyzing security events.

Integrating explainability into automated security workflows could significantly enhance intrusion response mechanisms.
Future work could explore how VoTE-XAI can be leveraged for mitigation, improving analyst decision-making, and reducing response times through user studies. Additionally, the incorporation of expert knowledge or real-time risk assessment to prioritize some (minimal) explanations over others could be further studied.

CF explanations provide insights into how a classification decision could change given minimal modifications to input features.
In the context of a 5G testbed, CF could help understand attack evolution. This is a topic we are currently pursuing further. By modifying key network attributes, we could infer how an attacker might evade detection, aiding in proactive defense. Evaluating the impact of adversarial perturbations can also help identify model weaknesses and enhance resilience against stealthy attacks.
    
\section*{Acknowledgment}
This work was carried out within the NEST project $AIR^2$, which is partially supported by the Wallenberg AI, Autonomous Systems and Software Program (WASP) funded by the Knut and Alice Wallenberg Foundation. It was also supported by ELLIIT, the Excellence Center at Linköping – Lund in Information Technology.
The authors would like to thank Elisa Bertino and Kazi Samin Mubasshir for collecting the MSA dataset and for providing valuable information during the project.

\bibliographystyle{elsarticle-num} 
\bibliography{reference.bib}

\end{document}